\documentclass[iop,apjl]{emulateapj}

\shorttitle{Discovery of molecular gas around HD\,131835}

\shortauthors{Mo\'or et al.}

\begin{document}


\title{Discovery of molecular gas around HD\,131835 in an APEX molecular line survey of bright debris disks}

\author{A. Mo\'or\altaffilmark{1}}
\email{moor@konkoly.hu}
\author{Th.~Henning\altaffilmark{2}}
\author{A. Juh\'asz\altaffilmark{3}}
\author{P. \'Abrah\'am\altaffilmark{1}}
\author{Z. Balog\altaffilmark{2}}
\author{\'A. K\'osp\'al\altaffilmark{1,2}}
\author{I. Pascucci\altaffilmark{4}}
\author{Gy.~M. Szab\'o\altaffilmark{1,5}}
\author{R. Vavrek\altaffilmark{6}}
\author{M. Cur\'e\altaffilmark{7}}
\author{T. Csengeri\altaffilmark{8}}
\author{C. Grady\altaffilmark{9,10}} 
\author{R. G\"usten\altaffilmark{8}}
\author{Cs. Kiss\altaffilmark{1}}

\altaffiltext{1}{Konkoly Observatory, Research Centre for Astronomy
  and Earth Sciences, Hungarian Academy of Sciences, PO Box 67, H-1525
  Budapest, Hungary}

\altaffiltext{2}{Max-Planck-Institut f\"ur Astronomie, K\"onigstuhl
  17, 69117 Heidelberg, Germany}

\altaffiltext{3}{Institute of Astronomy, Madingley Road, Cambridge CB3, OHA, UK}

\altaffiltext{4}{Lunar and Planetary Laboratory, University of Arizona, Tucson, 
AZ 85721, USA}

\altaffiltext{5}{ELTE Gothard Astrophysical Observatory, Szent Imre herceg 
\'ut 112, H-9700 Szombathely, Hungary}

\altaffiltext{6}{Herschel Science Centre, ESA/ESAC, PO Box 78,
  Villanueva de la Ca\~nada, 28691, Madrid, Spain}

\altaffiltext{7}{Instituto de F\'{i}sica y Astronom\'{i}a, Universidad de Valpara\'{i}so, Chile}

\altaffiltext{8}{Max-Planck-Institut f\"ur Radioastronomie, Auf dem
  H\"ugel 69, 53121 Bonn, Germany}

\altaffiltext{9}{NASA Goddard Space Flight Center, Code 667,
  Greenbelt, MD 20771, USA}

\altaffiltext{10}{Eureka Scientific, 2452 Delmer Street, Suite 100,
  Oakland, CA 94602, USA}


\begin{abstract}

Debris disks are considered to be gas-poor, but recent observations revealed molecular 
or atomic gas in several 10--40\,Myr old systems. We used the APEX and IRAM\,30m 
radiotelescopes 
to search for CO gas in 20 bright debris disks. In one
case, around the 16\,Myr old A-type star HD131835, we discovered a new gas-bearing debris disk, 
where the CO 3--2
transition was successfully detected. 
No other individual system exhibited a measurable CO signal.
Our {\sl Herschel Space Observatory} far-infrared images of HD\,131835 marginally resolved the disk 
both at 70 and 100{\micron},
with a characteristic radius of $\sim$170\,au. While in stellar properties HD\,131835 resembles 
$\beta$\,Pic, its dust disk
properties are similar to those of the most massive young debris disks. 
{With the detection of gas in HD\,131835 the number of known debris disks with 
CO content has increased to four, all of them encircling young ($\leq$40 Myr) A-type stars.}
Based on statistics within 125\,pc, we suggest that the presence of detectable 
amount of gas in the most massive
debris disks around young A-type stars is a common phenomenon. 
Our current data cannot conclude on the origin of gas in
HD\,131835. If the gas is secondary, arising from the disruption of planetesimals, 
then HD\,131835 is a comparably young and in
terms of its disk  more massive analogue of the $\beta$\,Pic system. 
However, it is also possible that this system
similarly to HD\,21997  possesses a hybrid disk, 
where the gas material is predominantly primordial, while the dust
grains are mostly derived from planetesimals.

\end{abstract}


\keywords{circumstellar matter --- infrared: stars ---  stars:
  individual (HD\,131835)}




\setcounter{footnote}{0}

\section{Introduction}
\label{intro}
Stars can be accompanied by circumstellar matter throughout their 
life. In the early phase of their evolution, they are surrounded by 
massive gas-rich primordial disks that form during the 
collapse of the molecular cloud core as a consequence of angular momentum 
conservation \citep[e.g.,][]{williams2011}. According to the current paradigm, most of these disks 
dissipate within a few million years \citep{zuckerman1995,pascucci2006}. 
During this evolution disk material accretes onto the star, forms planets, or is removed from 
the system through outflows and photoevaporation \citep[][and references therein]{alexander2014}. What left is a tenuous debris disk
which differs from their predecessors in many ways. Dust grains in these disks are second 
generation: collisional erosion and evaporation of previously formed planetesimals 
provides a continuous replenishment of dust particles removed by 
the stellar radiation or wind on a significantly shorter timescale 
than the age of the star \citep{backman1993,wyatt2008,krivov2010}. Debris disks 
are thought to be gas-poor with a 
significantly lower gas-to-dust ratio than that of primordial disks \citep[e.g.][]{matthews2014b},
 because processes like collisions, evaporation, and photodesorption from icy grains only produce a small 
 amount of secondary gas \citep{beust1994,cm2007,grigorieva2007,zuckerman2012}.

One way to study the dust component in debris disks is to observe its
thermal emission at infrared (IR) and millimeter wavelengths.
Thanks to a series infrared space missions now we know hundreds of 
debris disks. The spectral energy distribution (SED) of the disks' excess emission 
is routinely used to infer their fundamental properties such as the characteristic dust 
temperature and the amount of dust grains, while resolved images give information about
their spatial extent and even the grain size distribution \citep[e.g.][]{booth2013,morales2013,pawellek2014}. 
Due to the intimate link between the grains 
and larger parent planetesimals the study of dust disk properties can also give 
insight into the characteristics and evolution 
of parent planetesimal belt(s) \citep{wyatt2008}. 

Observing the signatures of presumably very small amounts of gas is a more challenging 
task. {Despite extensive surveys \citep{dent2005,dent2013,moor2011b,hales2014,rm2014,rigliaco2015}, 
the number of debris disks with known gas component is still very limited.}
In the edge-on disks around $\beta$\,Pic and HD\,32297 the gas was 
firstly detected via absorption lines \citep{slettebak1975,hobbs1985,redfield2007}. At 49 Cet and HD\,21997 a 
substantial amount of CO gas has been observed \citep{zuckerman1995,moor2011b}, 
while the gaseous disks around HD\,172555, HD\,181296 and AU\,Mic were identified based on 
\ion{O}{1}, \ion{C}{2} and fluorescent H$_2$ line emissions, respectively \citep{rm2012,rm2014,france2007}. 
{Recently, \citet{rigliaco2015} identified 
H\,I (7--6) lines in the mid-infrared Spitzer/IRS 
spectra of 8 young (5--20\,Myr old) F7-K0 type stars that 
hosted debris disks.  
It is yet an open question whether these emission 
lines are linked to low level accretional processes and thereby 
indicate the presence of long lived gaseous circumstellar 
disks or they have pure chromospheric origin \citep{rigliaco2015}.}

{While debris dust disks have been found around stars with a variety of ages
the current sample of gaseous disks is mostly limited to young (typically 10--40\,Myr old) stars.
}
The origin of gas in these systems is not unambiguous yet. 
Because of their relatively young age we cannot exclude that some of them 
harbor hybrid disks where the dust has secondary origin and produced from 
planetesimals, while the gas is predominantly primordial, the remnant of the 
original disk. Indeed, as one of the early discoveries of ALMA, our group 
identified such a hybrid disk around the $\sim$30-Myr-old HD\,21997 \citep{kospal2013}.
This discovery challenges the current paradigm on the timescale of primordial disk evolution. 
Nevertheless, based on this small and, in many ways, limited sample we cannot draw general and 
comprehensive conclusions about the production mechanism and long term  evolution of 
the gas component in debris disks.

Motivated to find additional examples of gaseous debris disks  
and scrutinize the fundamental properties of their gas and dust contents, we used the 
Atacama Pathfinder Experiment\footnote{This publication is based on data acquired with the Atacama Pathfinder Experiment 
(APEX). APEX is a collaboration between the Max-Planck-Institut fur Radioastronomie, 
the European Southern Observatory, and the Onsala Space Observatory.}  and 
the IRAM\,30m\footnote{Based on observations carried out with the IRAM 30m Telescope. 
IRAM is supported by INSU/CNRS (France), MPG (Germany) and IGN (Spain).} radio telescopes to 
search for molecular gas in 20 debris disks in the rotational lines of 
CO. This work is a continuation of our similar survey of 20 objects \citep{moor2011b}.
In this paper we review the
results of the new survey and report on the discovery of a gaseous debris 
disk around HD\,131835, a $\sim$16\,Myr old member of the Upper Centaurus Lupus 
association. For this object we present additional photometric and spectroscopic observations 
with the {\sl Herschel Space Observatory} \citep{pilbratt2010}.

\section{Observations and data reduction} \label{obsanddatared}

\subsection{Sample selection} \label{selection}
In our previous survey we searched for gas in 20 young ($<$50\,Myr) debris disks with high fractional 
luminosity ($L_{IR}/L_{bol} = f_d > 5\times10^{-4}$) around A- and F-type stars \citep{moor2011b}. 
The current project focuses on 20 additional debris disks.
We included seven disks where the selection criteria were similar to those
applied in the first survey, except that we allowed lower fractional 
luminosities down to  $f_d = 10^{-4}$. 
We added four targets (HD\,10939, HD\,17848, HD\,161868 and HD\,182681) 
which are similar to the gaseous 49\,Cet and HD\,21997 
systems in that they harbor cold, very extended debris disks. 
By including six debris disks around G--K-type star we also extended the sample  
to Sun-like stars. 
Recently, \citet{montgomery2012} and \citet{welsh2013} reported nightly variability  
of the Ca\,II K absorption line in the spectra of some A-type debris disk host stars, indicating the
presence of circumstellar gas.
Note that among the known gaseous disks, $\beta$\,Pic, 49\,Ceti, and HD\,172555 also show similar 
variable Ca absorption
\citep{ferlet1987,montgomery2012,kiefer2014}. 
Thus we selected three nearby ($<$75\,pc) targets, HD\,110411, HD\,182919, and HD\,183324, from 
this sample. 
Basic properties of the selected targets are summarized in Table~\ref{basictable}.

\subsection{CO line observations} \label{apexobs}

\paragraph{Observations with APEX.}
The majority of our observations were carried out with the 12\,m Atacama Pathfinder Experiment 
radio telescope \citep[APEX,][]{gusten2006} in the framework of three different programmes.
All of our targets were observed in the $J$=3$-$2 transition of $^{12}$CO at 
a rest frequency of $\nu$ = 345.796\,GHz. 
In program M-087.F-0001-2011 the observations were performed 
with the Swedish Heterodyne Facility Instrument/APEX2 \citep[SHeFI,][]{vassilev2008} receiver, while in
M-092.F-0012-2013  and M-093.F-0010-2014 
the First Light APEX Submillimeter Heterodyne receiver \citep[FLASH+,][]{klein2014} was used.
The latter is a dual-frequency receiver that operates simultaneously in the 345\,GHz and 
461\,GHz atmospheric windows, therefore 
we obtained $^{12}$CO $J$=4$-$3 line ($\nu$ = 461.041\,GHz) observations as well. 
For HD\,131835, CO (2--1) line observations were also conducted using the 
SHeFI/APEX1 instrument. 

For SHeFI observations acquired before June 2011 we utilized the 
Fast Fourier Transform Spectrometer (FFTS) backend with a spectral resolution of 488\,kHz 
(0.42\,km\,s$^{-1}$ at $J$=3--2 transition).
For later SHeFI measurements 
the eXtended bandwidth Fast Fourier Transform Spectrometer \citep[XFFTS,][]{klein2012} with 32768 
channels providing a spectral resolution of 77\,kHz (0.066\,kms$^{-1}$ at $J$=3--2 transition)
was connected to the receiver. FLASH+ was always connected to the XFFTS backend. 
Beam sizes of APEX are $\sim$27{\arcsec}, $\sim$18{\arcsec} and $\sim$14{\arcsec} at 230, 345 and 460\,GHz, respectively.
All of our APEX observations were performed in on-off observing mode.

The CO (3--2) spectral line observations of HD\,131835 were further supplemented by 
yet unpublished measurements obtained in our 
previous APEX programme E-083.C-0303 \citep[for details see,][]{moor2011b}.

\paragraph{Observations with the IRAM 30\,m telescope.}
The three targets selected because of variable Ca\,II K absorption line  
were observed with
the IRAM 30\,m telescope using the multi-band heterodyne Eight MIxer Receiver 
\citep[EMIR,][]{carter2012} as part of the IRAM programme No.~172-13. 
We searched for gas in our targets at 230.538\,GHz, the $J=2-1$ transition of the $^{12}$CO.
The observations were conducted in wobbler-switching on-off mode with a wobbler throw of 60{\arcsec}.
For the backend, we used the new Fast Fourier Transform Spectrometer (FTS)
with a frequency resolution of 200\,kHz providing a velocity resolution of 
$\sim$0.25\,km\,s$^{-1}$ in the $J$=2$-$1 transition.
Table~\ref{obslog} summarizes the main characteristics of the observing programmes.

\paragraph{Data reduction.} 
Both the APEX and IRAM spectra have been processed using the GILDAS/CLASS 
package\footnote{\url{http://iram.fr/IRAMFR/GILDAS/}}. 
For the final average spectrum, we discarded {noisy scans} 
and a baseline was subtracted from each individual scan. 
The baseline was typically linear, except for a few cases where we used second order polynomials.
The final spectrum was derived as an average of the individual spectra weighted by the inverse 
square of their rms noise.
As a final step the obtained antenna temperatures were converted to line flux densities. 
For the APEX data we used Kelvin-to-Jansky conversion factors of 39, 41, and 48\,Jy~K$^{-1}$ 
for CO (2--1), CO (3--2), and CO (4--3) transitions, respectively. For the IRAM CO (2--1) observations
the corresponding number was 7.8\,Jy~K$^{-1}$.
These conversion factors
were taken from 
the relevant APEX\footnote{http://www.apex-telescope.org/telescope/efficiency/} 
and IRAM\footnote{http://www.iram.es/IRAMES/mainWiki/Iram30mEfficiencies} 
home pages.

\paragraph{Outcome of the survey.} \label{outcome}

\begin{figure} 
\includegraphics[scale=.50,angle=0]{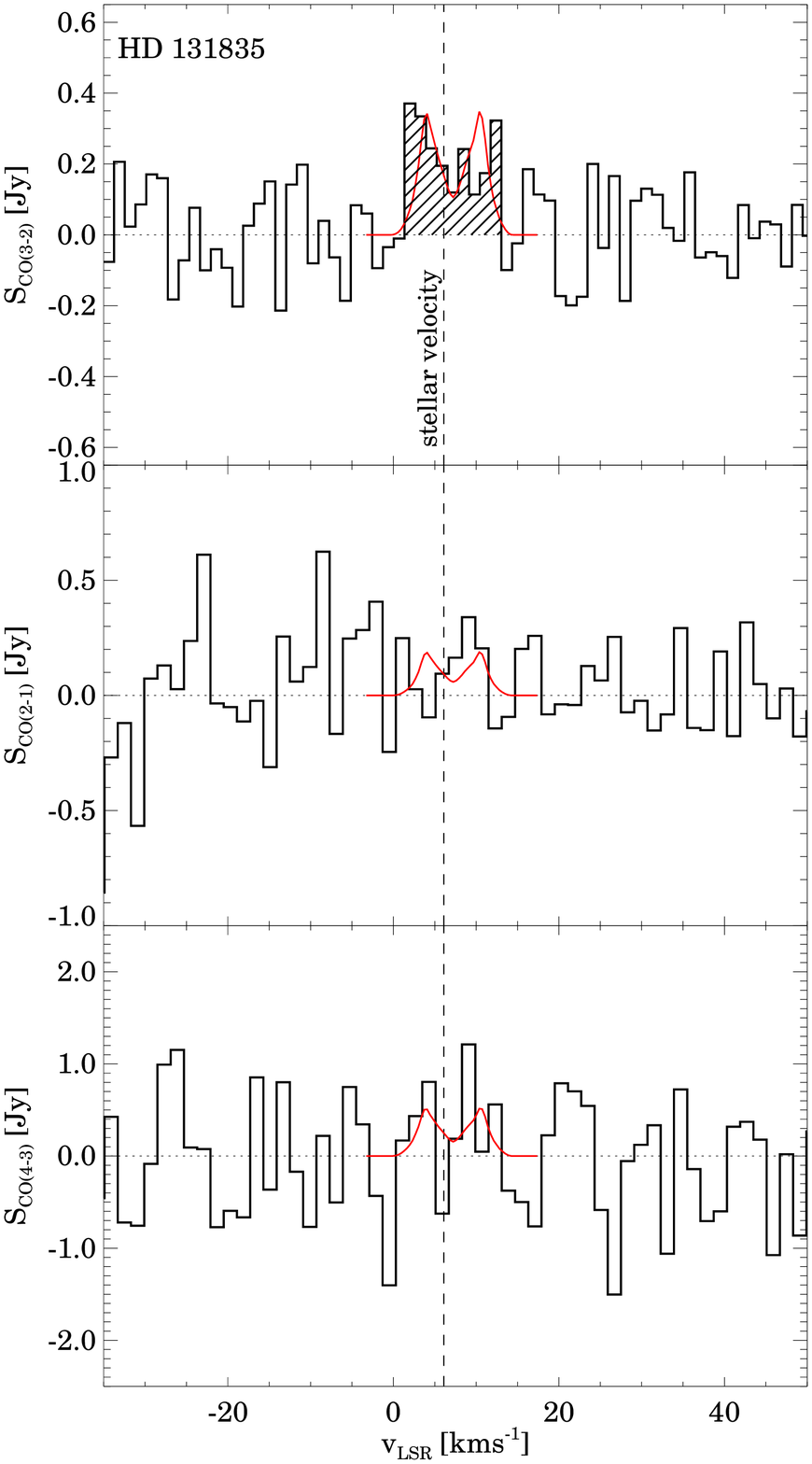}
\caption{ {\sl APEX CO(3--2), CO(2--1), and CO(4--3) spectra of HD\,131835. 
The dashed line marks the radial velocity of the star. In the upper panel 
the hatched area shows the part of the spectrum considered in the line 
flux integration (Sect.~\ref{apexobs}). The best fit model spectra (see Sect.~\ref{gasparams}) are  plotted by red lines.
}
\label{co32}
}
\end{figure}

CO emission was evident only for HD\,131835, where the integrated line flux 
of  the 3--2 transition was successfully 
detected at 
a 5.0$\sigma$ level. Figure~\ref{co32} shows the baseline subtracted CO (3--2) profile of this target. 
The measured line has a width of roughly 12\,km~s$^{-1}$ and  
is centred at a velocity of 7.2\,km~s$^{-1}$ with respect to the Local Standard of Rest (LSR), 
in very good agreement with the systemic LSR velocity of the star (6.1$\pm$1.1\,km~s$^{-1}$, computed from the radial 
velocity derived in Sect.~\ref{ferosanalysis}). 
The measured line tentatively shows a double peak profile in accordance what we expect for a gas disk 
in Keplerian rotation around a star. 
The peak flux is $\sim$370\,mJy, the integrated line flux -- obtained by integrating the line over a 
12\,km~s$^{-1}$ interval -- is 2.74$\pm$0.55\,Jy~km~s$^{-1}$. 
By inspecting the same velocity interval in the $J$=2--1 and 4--3 spectra (Fig.~\ref{co32})
we found no statistically significant evidence for lines, the obtained integrated line 
fluxes are 1.60$\pm$0.78 and 4.46$\pm$2.99\,Jy~km~s$^{-1}$, respectively. 
We note that HD 131835 was previously observed in the CO 2--1 transition by \citet{kastner2010} 
resulting in a non-detection. This observation was about two times less sensitive than ours. Assuming 
 a linewidth of 5\,km\,s$^{-1}$, \citet{zuckerman2012} deduced a 5$\sigma$ line flux upper limit 
of 5.1\,Jy~km~s$^{-1}$ from this measurement, that is compatible with our result.

None of our other targets were detected at any of the CO transitions.
For these sources, 
upper limits were computed as $S_{rms} \Delta v \sqrt{N}$ 
where $S_{\rm rms}$ is the 1\,$\sigma$
measured noise, $\Delta v$ is the velocity channel width, and $N$ is 
the number of velocity channels over
an interval of 10\,km\,s$^{-1}$.
The obtained line fluxes and upper limits are listed in Table~\ref{basictable}.

\subsection{Additional observations for HD~131835} \label{addobs}

Because of the detection of CO (Sect.~\ref{outcome}),
for HD\,131835 we performed several additional observations.  
In order to better characterize its spectral energy distribution at far infrared/submillimeter wavelengths 
and to search for \ion{O}{1} and \ion{C}{2} emission from the disk we obtained photometric and spectroscopic 
measurements using the Photodetector Array Camera and Spectrometer \citep[PACS,][]{poglitsch2010} and 
the Spectral and Photometric Imaging Receiver \citep[SPIRE,][]{griffin2010} 
onboard the {\sl Herschel Space Observatory} (program id OT2\_amoor\_3). 
These data were complemented by mid-infrared spectra 
obtained with the {\sl Spitzer Space Telescope} \citep{werner2004}. 
To investigate the stellar properties 
a high-resolution ground-based optical spectrum was also taken.

\subsubsection{{\sl Herschel} Observations} \label{herschelobs}

\paragraph{PACS maps.}
Far-infrared maps were obtained with the PACS detector on 2012 September 9 and 11 in mini scan-map mode 
(PACS Observer's Manual v2.5\footnote{http://herschel.esac.esa.int/Docs/PACS/html/pacs\_om.html})
using a scan speed of 20{\arcsec}\,s$^{-1}$ with 10 scan-legs of 3{\arcmin} length 
separated by 4{\arcsec}. We made two scans with scan angles of 70{$\degr$} and 110{$\degr$} 
both at 70\,{\micron} (OBSID: 1342250784, 1342250785) and at 
100\,{\micron} (OBSID: 1342250868, 1342250869).  
Since the PACS photometer observed in two bands simultaneously 
(at 160{\micron} in addition to 70{\micron} or 
100{\micron}) these
measurements provided four 160{\micron} scans as well.

Data processing was done with the Herschel Interactive Processing Environment 
\citep[HIPE,][]{ott2010} version 13 using PACS calibration tree No.\,65 and 
the standard HIPE script optimized for the reduction of bright sources. 
 Additionally we used the recently developed "gyro" correction 
to reduce the pointing jitter.
We selected only those data frames from the timeline  where the actual scan speed of the spacecraft 
was between 15 and 25{\arcsec}~s$^{-1}$.
To eliminate the marked low-frequency (1/f) noise we  
applied highpass filtering with filter width parameters of 15, 20, and 35 for
the 70, 100, and 160{\micron} data, respectively. 
In order to avoid flux loss the 25{\arcsec} radius vicinity of our targets 
was excluded from the filtering. 
For glitch removal we used the second-level deglitching algorithm.
As a final step, in each band we applied the {\sc 'photProject'} task to combine all frames
into a map using the default pixel fraction (1.0) and pixel sizes of 1\farcs1, 1\farcs4, and 
2\farcs1 at 70, 100, and 160\,{\micron}, respectively.

HD\,131835 was clearly detected in all PACS bands.  
We performed aperture photometry on the source using a radius of 18{\arcsec}, while the sky 
background was estimated in an annulus between 40{\arcsec} and 50{\arcsec}. The aperture was placed 
at the source's centroid position. The offsets between the derived centroids and 
the targets' optical position (corrected for the proper motion using the 
epochs of PACS observations) were $<$1{\arcsec} in all bands. 
For sky noise determination, we distributed sixteen apertures with radii of 18{\arcsec} (identical to the
the source aperture) randomly along
the background annulus. By performing aperture photometry without background subtraction 
in each aperture, we computed the sky noise as the standard deviation 
of these background flux values.
We applied aperture correction
in each band, by using correction 
factors taken from the calibration file attached to the measurement. 
To derive the final uncertainty of our photometry we added the 
measurement errors and the calibration uncertainty \citep[7\%,][]{balog2013} quadratically.
The obtained flux density values and their uncertainties are listed in Table~\ref{phottable}.

\paragraph{PACS spectroscopy.}
We carried out PACS spectroscopic observations 
centred on the \ion{O}{1} 63{\micron} (OBSID: 1342248686) and the \ion{C}{2} 
158{\micron} (OBSID: 1342248687) lines.
These fine structure lines if arising from circumstellar disks are unresolved 
in {\sl Herschel} PACS observations.
Observations were performed on 2012 July 29 using
the Line Spectroscopy observing template that enables the 
coverage of a small wavelength interval around these lines. 
To efficiently eliminate telescope and sky background, 
the measurements were obtained in chop-nod mode.
The PACS spectrometer array consists of 5$\times$5 spectral pixels (spaxel), each  
having a size of 9\farcs4. We made small 3$\times$3 and 2$\times$2 raster maps 
with raster point/line step 
sizes of 3{\arcsec} and 2{\arcsec} for the \ion{O}{1} and \ion{C}{2} observations, respectively.
The maps were centred on the target, and they were repeated 
two and four times in the case of \ion{O}{1} and \ion{C}{2} observations, respectively.

Raster map observing mode has been selected to mitigate the risk of
eventual pointing offsets of the telescope significant larger than
2{\arcsec}. While the 47{\arcsec}$\times$47{\arcsec} spatial footprint of the integral field
unit is sufficiently large to almost entirely cover the spectrometer
spatial beam, a single spaxel of 9.4{\arcsec}$\times$9.4{\arcsec} size is strongly
undersampling the beam especially at the short wavelength of the \ion{O}{1}
line. As a consequence of undersampling, in the baseline pointed
observing scheme the spectrum of a point source is extracted from the
central spaxel of the 5$\times$5 unit applying a correction for fluxes due
to small pointing offsets from the centre of the central spaxel and
point jitter. However, if the source flux is lower than about a few
Jy per spaxel then correction factors estimated from the source
continuum sampled by the eight neighbors of the central spaxel may
become unreliable and the spectrum extracted from the central spaxel
may have to be left uncorrected for mispointing.

To overcome this problem, we apply mapping observation of the same
field of view with sub-spaxel raster step size. When data from raster
positions is combined the sampling of the beam is highly improved
even in case of an eventual pointing offset of the raster central
position. This technique directly provides a high resolution image of
the source where the spectrum can be extracted within a synthetic
aperture centered on the measured peak position.

We apply the telescope normalization scheme for spectro-photometric
flux calibration, where the Herschel telescope background calibrated
on Neptune is used as an absolute radiometric reference. The method
efficiently eliminates any drifts with time in the system response at
the frequency of individual chopper cycles resulting the offset
signals in nod A and nod B positions perfectly cancel each other.
This method may improve signal-to-noise - especially for faint
sources in the sub-Jansky per spaxel regime - comparing to the
baseline flux calibration method which relies on response estimates
from calibration blocks executed at the beginning of the observation
and propagated to the covered wavelength range through the relative
spectral response function.

Flux calibrated IFU cubes obtained at subsequent raster positions are
combined into a single spectral cube using a 3D drizzling algorithm \citep{regibo2012}
available in the PACS interactive pipeline in HIPE. The strength of
the drizzling algorithm is that the convolution with the detector
footprint is minimized especially when we apply the "gyro"
improved high spatial resolution reconstruction of instantaneous
pointing jitter. The projected pixel size is adjusted to Nyquist
sample the beam at the line central wavelength.

In the last data processing step, spectra from the drizzled spectral
cube are combined within an extraction aperture of 6.3{\arcsec} radius
centered on the source position.

\begin{figure} 
\includegraphics[scale=.5,angle=0]{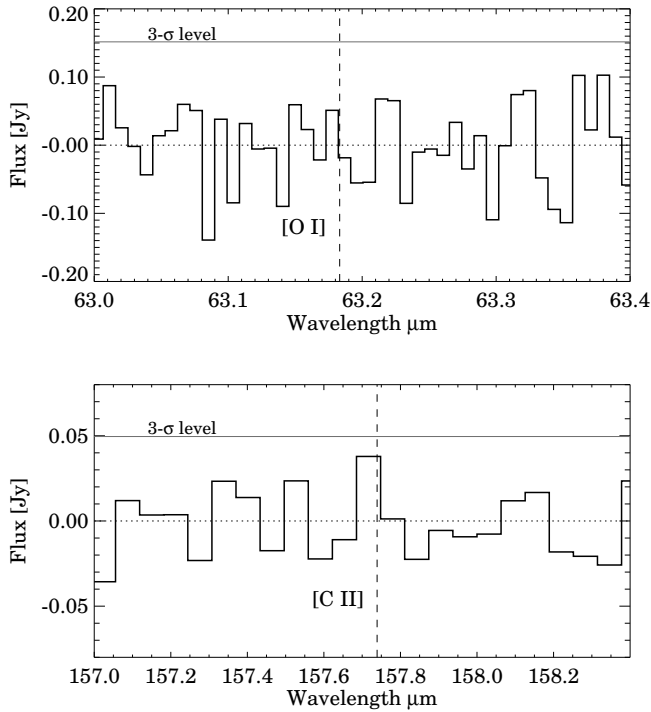}
\caption{ {\sl Continuum subtracted spectra of HD\,131835 at 63{\micron} (upper panel) and 157{\micron} (lower panel). }
\label{herschellines}
}
\end{figure}

Figure~\ref{herschellines} shows the obtained \ion{O}{1} and \ion{C}{2} spectra.
Though in the \ion{C}{2} spectrum there is a peak at 157.7{\micron} its significance is only 
2.3\,$\sigma$, thus
neither the \ion{O}{1} or \ion{C}{2} line was detected towards our target.
In order to derive upper limits for the line fluxes we used the continuum 
subtracted spectra and determined the rms noise ($\sigma_{\rm rms}$) in a 20 pixels wide window 
centred on the expected line wavelength. 
The upper limits were computed 
as $3 \times \sigma_{\rm rms} \times \Delta \nu \times \sqrt{p},$
 where $\Delta \nu$ is the width of 
one pixel in Hz, while $p$ is the width of an unresolved emission line  
in pixels. Values of $p$ were calculated as the ratio of the full width 
half maximum (FWHM) of unresolved lines (taken from PACS Observer's Manual v2.5, 
0.020{\micron} and 0.126{\micron} for \ion{O}{1} and \ion{C}{2} spectra, respectively) 
to the appropriate pixel widths (in {\micron}).
The obtained upper limits for the \ion{O}{1} and \ion{C}{2} 
lines are 1.5$\times$10$^{-18}$\,Wm$^{-2}$ and 5.3$\times$10$^{-19}$\,Wm$^{-2}$, respectively.

\paragraph{SPIRE maps.}
We observed HD\,131835 with the SPIRE Photometer on 2012 August
9, in Small Scan Map mode (Spire Handbook 
v2.5\footnote{http://herschel.esac.esa.int/Docs/SPIRE/spire\_handbook.pdf}) 
at 250, 350, and 500\,{\micron} simultaneously. Data were processed with HIPE 13 
following the standard pipeline processing steps with default values for all applied tasks. 
The final maps were produced using the {\sc 'naiveMapper'} task.
The beam sizes are 17\farcs6, 23\farcs9, and 35\farcs2 at 250, 350, and 
500\,{\micron}, respectively, and the maps were resampled to pixel sizes of 
6{\arcsec}, 10{\arcsec}, and 14{\arcsec} at these wavelengths.

Our target is detected in all bands. 
We used point-spread function (PSF) photometry to extract 
the source fluxes, the corresponding SPIRE beam profiles were taken from the 
calibration context. 
This method yielded flux densities of 156.4$\pm$7.6\,mJy, 
84.3$\pm$6.9\,mJy, and 35.4$\pm$8.6\,mJy at 250, 350, and 500\,{\micron}, 
respectively. 
In order to validate our flux values the photometry was also 
performed using the SPIRE Timeline Fitter task in HIPE. Instead of using 
the final maps, this task fits two dimensional elliptical or circular 
Gaussian functions to the baseline-subtracted timeline data of the photometer 
at the coordinates of the source \citep[][]{bendo2013}.
The radius of the region that includes the peak
of the source was set to 22{\arcsec}, 30{\arcsec}, 42{\arcsec} for 250, 350, and 500\,{\micron} maps.
The background was estimated from an annulus between 300{\arcsec} and 350{\arcsec} centred 
on the source. At 250 and 350\,{\micron} we used elliptical Gaussian 
during the fitting, at 500\,{\micron}, because of the source's lower signal-to-noise ratio, 
we applied a circular Gaussian. 
Using the Timeline Fitter we obtained flux densities of 165.4$\pm$4.4\,mJy, 
77.8$\pm$4.4\,mJy, and 44.1$\pm$8.3\,mJy at 250, 350, and 500\,{\micron}, 
respectively, in good accordance with the ones derived 
from PSF photometry. 
In the following analysis, we will use the PSF photometry.
The final uncertainties
were derived as the quadratic sum of the measurement errors and the
overall calibration uncertainty of the SPIRE photometer \citep[5.5\%][]{bendo2013}.
The derived results are quoted in Table~\ref{phottable}.

\subsubsection{Spitzer/IRS observation} \label{irsspectra}
HD\,131835 was observed with the InfraRed Spectrograph \citep[IRS,][]{houck2004} 
twice, first on 2007 August 30 (PID40651), then on 2008 April 1
(PID40235). Both spectra were obtained in Staring mode using 
the low-resolution IRS modules (SL and LL), covering the $5.2-38\,\mu$m wavelength 
range with a spectral resolution of $R=60-120$. We retrieved the processed spectra 
from the CASSIS\footnote{The Cornell Atlas of Spitzer/IRS Sources (CASSIS) is a 
product of the Infrared Science Center at Cornell University, supported by NASA and 
JPL.} database \citep{lebouteiller2011}. 
As post-processing, some outlying data points were discarded.
We fitted polynomials to the data of individual IRS modules using a robust method and 
then searched for data points outlying by more than 4$\sigma$ from these fits. 
We found that there is a mismatch between the SL1 and LL2 modules in both spectra.
Since the shortest wavelength parts of the spectra match perfectly with the predicted 
photospheric fluxes (see Sect.~\ref{stellarprops}), the modules were stitched 
together by scaling the LL modules to the SL ones. 
The multiplicative scaling factors were estimated from the overlapping 
spectral regions of the SL1 and LL2 modules. 

The spectra of HD\,131835 do not show any prominent silicate features.
{Following \citet{redfield2007b} and \citet{rigliaco2015}, we also looked for atomic metal and hydrogen lines,  
and rotational lines of molecular hydrogen in the spectrum.} 
However, none of the lines are detected.
We compared the two IRS spectra of HD\,131835 obtained at
different epochs. The flux differences were typically in the 
order of 5\% over the whole spectral range, i.e.  
the continuum level was unchanged. 
For the further analysis the two spectra were combined using a 
simple weighted average.

For the SED modelling process 
the combined IRS spectrum was split into 15 adjacent
wavelength bins between 6 and 35\,{\micron}. 
The derived flux density values are listed in Table~\ref{phottable}.
Their uncertainties were computed by adding an 5\% absolute calibration 
uncertainty quadratically to the measured ones.

\subsubsection{FEROS observation}
High-resolution optical spectroscopy of HD\,131835 was performed using the
Fiber-fed Extended Range Optical Spectrograph
\citep[FEROS,][]{kaufer1999} installed on the 2.2\,m MPG/ESO telescope
in La Silla, on 2011 April 17 using the 
 ``object-sky'' configuration, with one fiber positioned at the target, and the other one on
the sky. The integration time was 180~s. FEROS has a mean resolving power of 
R $\sim$ 48000 and covers the wavelength range between 3500 and 9200 \AA{} 
in 39 echelle orders.
Data reduction, including
bias subtraction, flat-field correction, background subtraction, the
definition and extraction of orders, and wavelength calibration, was
performed using the FEROS Data Reduction System pipeline at the
telescope.

\section{Results} \label{results}

\subsection{Stellar properties of HD\,131835} \label{stellarprops} 

HD\,131835 is an A2-type star \citep{houk1982} located at 122.7$^{+16.2}_{-12.8}$\,pc away \citep{vanleeuwen2007}.
Based on astrometric data made with ESA's {\sl Hipparcos} satellite, \citet{dezeeuw1999} proposed 
the star to be a member of the $\sim$16\,Myr old \citep{pecaut2011} Upper Centaurus Lupus (UCL) subgroup of 
the Sco-Cen association 
with a membership probability of 95\%. According to our current knowledge HD\,131835 has no 
stellar companion \citep{kouwenhoven2005,wahhaj2013}. 

We used spectroscopic and photometric data to model the stellar photosphere and to estimate the fundamental 
stellar properties of HD\,131835. Moreover, supplementing the {\sl Hipparcos} astrometric data 
by our ground-based radial 
velocity measurement, its membership in UCL was also reconsidered.

\paragraph{Analysis of the FEROS spectrum.} \label{ferosanalysis}   
In order to estimate the radial velocity and different 
stellar parameters of our target the measured FEROS spectrum 
was compared with templates taken from the spectral library 
of \citet{munari2005}. In this spectral library the continuum 
normalization was performed by dividing the absolute flux 
spectrum by its calculated continuum. Because of the applied 
method, in some spectral regions the continuum of the model 
spectra can significantly differ from a continuum that would 
be obtained using a regular normalization scheme of observational 
spectroscopy. Therefore first we used high order polynomials to 
flatten the model spectra, and then the same method was applied 
in the continuum normalization of our observed spectrum as well.
We used an iterative method to estimate the stellar parameters.    
We first applied the cross-correlation technique to derive the radial velocity 
of the star by convolving the measured spectrum with a template 
taken from the spectral library. We selected a template spectrum 
with $[Fe/H] = 0.0$, $\log{g} = 4.0$, and $T_{\rm eff} = 8750$\,K 
(corresponding to the A2 spectral type of the target quoted by the 
SIMBAD). 
The cross correlation function was calculated by the {\sc fxcor} task in
IRAF. After transforming the measured spectrum to the laboratory system we 
compared it with a grid of Munari synthetic spectra, using the
4000--6200\AA\ wavelength range, excluding the $H_\beta$ region and
Na~D lines. The grid was compiled by varying 
the effective temperature, surface gravity, and projected rotational velocity 
of the model spectra, while the metallicity was fixed to $[Fe/H] = 0.0$.    
By finding the global minimum we used the derived parameters to 
repeat the radial velocity determination and then recompute the stellar parameters. 
This process yielded the following best-fit parameters: $T_{\rm eff} =
8250\pm250$K, $\log\ g = 4.0\pm0.5$, $v\sin\ i = 100\pm15$kms$^{-1}$. 
Based on dwarf spectral type vs. effective temperature scale presented in 
\citet{pecaut2013} 
the derived $T_{\rm eff}$ rather corresponds to a spectral type of A4.
The radial velocity was found to be 1.6$\pm$1.4\,km~s$^{-1}$, consistently 
with our previous measurement of 3.3$\pm$1.7\,km~s$^{-1}$ \citep{moor2006}. 
Taking the weighted average of the two radial velocity values we obtained 
$v_{\rm r} = 2.3\pm1.1$\,km~s$^{-1}$.

\begin{figure} 
\includegraphics[scale=.45,angle=0]{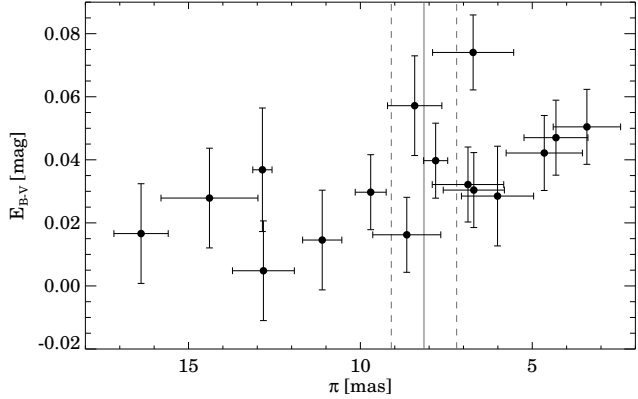}
\caption{ {\sl Derived reddening values as a function of trigonometric parallaxes 
taken from the Hipparcos catalogue 
for 16 stars located within 3{\degr} from HD\,131835. The solid and dashed vertical
lines show the measured parallax and its uncertainty for HD\,131835.}
\label{ebv}
}
\end{figure}

\paragraph{Analysis of photometric and near-IR data.}
To provide an additional independent estimate for fundamental stellar properties and model 
the stellar photosphere we fitted an ATLAS9 atmosphere model 
 \citep{castelli2003} to the optical and near-IR observations of the target.
Photometric data were taken from the {\sl Tycho~2} \citep{hog2000}, 
 {\sl Hipparcos} \citep{perryman1997}, and Two Micron All Sky Survey catalogues 
 \citep[2MASS,][]{cutri2003}. These data were further supplemented by Wide-field Infrared Survey Explorer 
 ({\sl WISE}) $W1$ band  
photometry {at} 3.4{\micron} from the {\sl WISE} All-Sky Database \citep{wright2010} and Johnson B, V photometry 
from the catalogue of \citet{slawson1992}.  
With a distance of $\sim$123\,pc, HD\,131835 could be outside the Local Bubble, thus 
its reddening might not be negligible. In order to provide a rough estimate of the reddening at the distance of 
HD\,131835, we collected those stars in its vicinity (with separations $<$3{\degr}) that have both {\sl Hipparcos}-based 
trigonometric parallax and measured Str\"omgren colour indices and H$\beta$ index 
in the photometric catalogue of \citet{hauck1998}. For these stars 
we derived $E(B-V)$ colour excesses (as $E(B-V) = E(b-y) / 0.74$) from the Str\"omgren data by using 
the appropriate calibration processes \citep{crawford1975,crawford1979,olsen1984}.     
Figure~\ref{ebv} shows the derived $E(B-V)$ values as a function of trigonometric parallaxes. 
At the distance of HD\,131835 (marked by a vertical line) the reddening is between $\sim$0.015 and 
$\sim$0.075\,mag. 
We note that consistent with our results, the extinction map of 
\citet{schlafly2011} shows a reddening of $\sim$0.083\,mag in the direction 
of our target. This value gives the total  
reddening within the Milky Way for the line of sight, thus it can be considered as an upper limit.        
Therefore in the photosphere modelling we fitted both the effective temperature and 
the reddening, the latter was limited between 0.0 and 0.08\,mag). 
By adopting solar metallicity
and $\log{g} = 4.25,$ our $\chi^2$ minimization yield{ed} $T_{\rm eff}$ =
8250$^{+250}_{-100}$\,K and $E(B-V)$ = 0.025$^{+0.05}_{-0.02}$. 
Considering these data and the ${\sl Hipparcos}$-based trigonometric distance 
we derived a luminosity of $L_{\rm bol}$ = 9.2$\pm$2.6\,L$_\odot$ for HD\,131835. 

\begin{figure} 
\includegraphics[scale=.45,angle=0]{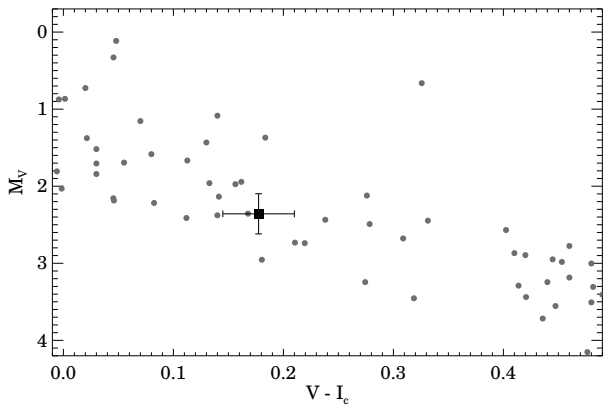}
\caption{ {\sl Absolute V band magnitude versus V$-$I$_{\rm c}$ 
color indices for known members of UCL \citep[the member list was taken from][]{dezeeuw1999} 
located within 150\,pc and having parallax error $<$10\% (gray dots) 
and for HD\,131835 (black square). 
Photometric data were taken from the {\sl Hipparcos} catalog and dereddened 
using extinction values from \citet{chen2011,chen2012} and, for HD\,131835, based on our result.}
\label{cmd}
}
\end{figure}

\paragraph{UCL membership.}
Using the new combined radial velocity and {\sl Hipparcos} astrometric
data of HD\,131835 we computed a Galactic space
motion of U\,=\,$-$5.8$\pm$1.3\,kms$^{-1}$, V\,=\,$-$18.4$\pm$2.1\,kms$^{-1}$, and W\,=\,$-$4.8$\pm$0.8\,kms$^{-1}$ 
with respect to the Sun. This space velocity is in perfect agreement 
with the charateristic space motion of UCL 
\citep[U\,=\,$-$5.1$\pm$0.6\,kms$^{-1}$, V\,=\,$-$19.7$\pm$0.4\,kms$^{-1}$, W\,=\,$-$4.6$\pm$0.3\,kms$^{-1}$,][]{chen2011}. 
The position of HD\,131835 in the color-magnitude diagram (Fig.~\ref{cmd}) matches well
 the locus defined by UCL members, indicating that its isochrone age
is consistent with that of the cluster. Our results confirm 
the previously proposed membership of HD\,131835 in UCL, therefore 
we adopted the age of the cluster, $\sim$16\,Myr, for our star.
With this age and our previous effective temperature and stellar luminosity estimates
we derived a stellar mass of 1.77$\pm$0.08\,M$_\odot$
by using solar metallicity isochrones from \citet{siess2000}.

\subsection{The dust disk around HD\,131835} \label{dustprops}

\begin{figure} 
\includegraphics[scale=.45,angle=0]{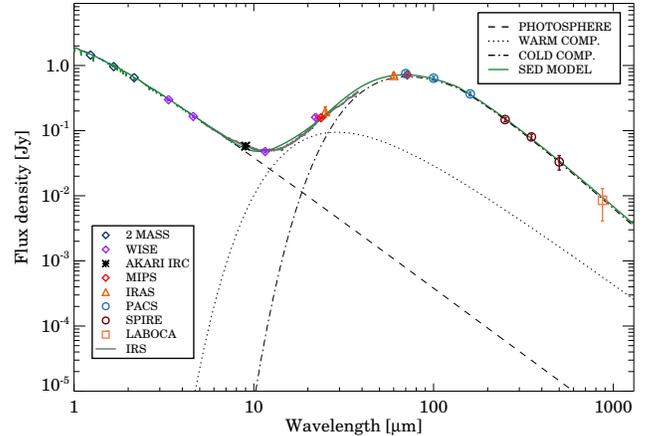}
\caption{ {\sl Color corrected SED of HD\,131835 overplotted by the fitted simple two-component 
model.}
\label{sed}
}
\end{figure}

\paragraph{SED modelling.} 
In order to construct the spectral energy distribution of 
HD\,131835, the previously mentioned PACS, SPIRE, and IRS data were 
further supplemented by infrared and submillimeter photometry from
the literature (Table~\ref{phottable}). The compiled SED is plotted in 
Figure~\ref{sed}. 
HD\,131835 exhibits strong excess at far-IR wavelengths, 
but as the IRS spectrum shows, the SED of the source already
starts to departing from photospheric emission at $\sim$7{\micron}, indicating 
the presence of warmer circumstellar dust as well. 
We used the ATLAS9 atmosphere model of HD\,131835 (Sect.~\ref{stellarprops}) to predict 
photospheric flux contribution at different mid- and far-IR wavelengths.
The average accuracy of the predicted photospheric fluxes is around 5\%.

The excess emission of debris disks is generally well fitted 
by a single (modified) blackbody or a combination of two different temperature 
blackbody components \citep{chen2014,kennedy2014}. 
This simple model can provide estimates of 
fundamental disk properties such as the characteristic dust temperature(s) and 
the fractional luminosity ($f_{\rm d}$ is the ratio of the luminosity
of the dust emission to the bolometric luminosity of the host star).

We fitted the excess emission of HD\,131835 by single- and two-temperature models.
For the single-temperature fitting we used a modified blackbody: 
\begin{equation}
F_{\rm \nu,exc} = A B_{\nu}(T_d) X_{\lambda}^{-1},
\end{equation}
where $F_{\rm \nu,exc}$ is the measured excess emission at $\nu$, $B_{\nu}$ is the Planck function, 
$T_{\rm d}$ is the dust temperature, $A$ is a scaling factor that is proportional to the 
solid angle of the emitting
region, while 
 $X_{\lambda} = 1$ if $\lambda \leq \lambda_0$ and  $X_{\lambda} = (\frac{\lambda}{\lambda_0})^\beta$ if $\lambda > \lambda_0$.  
In the two-temperature model we used a combination of a warmer simple blackbody (there is no sense to use 
a modified blackbody since the values of $\lambda_0$ and $\beta$ cannot be constrained well from the SED) 
and a modified blackbody.  
We applied a Levenberg-Marquardt algorithm \citep{markwardt2009} to find the 
best-fitting model. 
An iterative way was used to compute and apply color corrections for the photometric data 
during the fitting process \citep[see e.g.][]{moor2006}.
We found that only the two-temperature model (plotted in Figure~\ref{sed}) could reasonably fit both 
the mid- and far-IR/submm excess emission. 
This model provides dust temperatures of $T_{\rm d,w} = 176\pm22 $\,K and $T_{\rm d,c} = 71\pm3$\,K for the warm and cold dust 
components, respectively. We obtained $\lambda_0 = 136\pm30$\,{\micron} and $\beta = 0.50\pm0.16$ for the cold belt.
The warm component has a fractional luminosity of (8.2$\pm$1.9)$\times$10$^{-4}$ , while the cold component has a 
fractional luminosity of (2.2$\pm$0.2)$\times$10$^{-3}$.

We estimated the dust mass in the disk 
based on the submillimeter brightness
using the standard formula: $$
M_{d} = \frac{F_{\nu, \rm excess} d^2}{ B_{\nu}(T_{\rm d,c}) \kappa_{\nu}},$$
where $F_{\nu,{\rm excess}}$ is the measured excess at 500\,{\micron}
(we used this data point because the longest wavelength LABOCA measurement has very low 
signal-to-noise ratio), $d$ is the
distance to the source, $\kappa_{\nu} = \kappa_{0}
(\frac{\nu}{\nu_0})^{\beta}$ is the mass absorption coefficient, and
$B_{\nu}$ is the Planck function. 
By adopting $\kappa_{0} = $2\,cm$^2$\,g$^{-1}$ at $\nu_0 = 345$~GHz \citep[e.g.][]{nilsson2010}, and taking
$\beta$ and $T_{\rm d,c}$ values 
from the previous SED modelling, we derived a dust mass of 0.47$\pm$0.18\,M$_\oplus$ for the disk.

\paragraph{Spatial extent.}

In order to evaluate whether the disk is spatially extended its profiles measured at 70, 100, and 160\,{\micron}  
were compared with appropriate PSFs of the PACS instrument. The PSF profiles were compiled   
using mini-scan map observations of four stars ($\alpha$~Boo, $\alpha$~Tau, $\alpha$~Cet, $\beta$~And) 
that do not exhibit infrared excesses and served as fiducial standards in the calibration of PACS 
photometry \citep{balog2013}. 
For the data processing of these observations we used identical reduction steps as in the case of 
HD\,131835, and finally the obtained PSFs were rotated to match the roll angle of the telescope at
the time of observing our target. By fitting a two dimensional Gaussian function to the PSF 
profiles we derive averaged FWHMs of 5\farcs70$\pm$0\farcs02$\times$5\farcs59$\pm$0\farcs03 and 
6\farcs86$\pm$0\farcs04$\times$6\farcs76$\pm$0\farcs03 at 70 and 100{\micron}, 
respectively, while for the target we obtained 6\farcs49$\pm$0\farcs20$\times$5\farcs87$\pm$0\farcs19 and 
7\farcs74$\pm$0\farcs34$\times$7\farcs19$\pm$0\farcs32, implying 
that the disk is slightly elongated and marginally extended along its major axis 
at these wavelengths. At the longest wavelength the 
target's profile was consistent with that of the PSF measurements.
In order to derive the characteristic size, inclination, and position angle of the disk,
the 70 and 100\,{\micron} PACS images were fitted using a simple,
non-physical disk model grid in the same way as described in 
\citet{moor2015}. In this model we assumed that the dust emitting at these wavelengths 
is located in a narrow outer ring 
around the central star. 
The model has three free parameters, the average radius ($R_{\rm avg}$), the position angle ($PA$) and the inclination ($i$) 
of the disk, while the width of the disk was fixed to 0.1$R_{\rm avg}$ following 
\citet{booth2013}.
We used a Bayesian analysis in the selection of the best fitting model.
Our best solution has $R_{\rm avg} = 160\pm20$\,au, $PA = 54\pm$8{\degr}, and $i = 86^{+4}_{-10}${\degr} at 
70\,{\micron} and $R_{\rm avg} = 207\pm40$\,au, $PA = 39\pm$12{\degr}, and $i = 66\pm$24{\degr} at 100\,{\micron}.
The final disk parameters computed as a weighted average of those derived in the two bands are 
$R_{\rm avg} = 169\pm18$\,au, $PA = 49\pm$7{\degr}, and $i = 83^{+7}_{-9}${\degr}.
It is worth noting that the measured profiles can also be fitted
using a more extended disk model with a smaller inner radius and
a larger outer radius. 
Such a model, would provide
very similar $PA$ and inclination parameters as above, and the
average disk radius would also not change significantly.

The disk around HD\,131835 has also been successfully resolved at 11.7 and 18.3\,{\micron} using the 
Gemini South telescope \citep{hung2015} deriving a $PA$ of $\sim$61{\degr}
and an inclination of $\sim$74{\degr}. These parameters are broadly consistent 
with our results. 
By modeling the SED and the mid-IR images simultaneously, \citet{hung2015} proposed a combination of an
extended continuous power-law disk between 35 and 310\,au and two narrow rings at 105$\pm$5 and 220$\pm$40\,au 
stellocentric distances as the best-fitting solution. The three disk components are made of three different 
grain populations. All of the emitting grains are proposed to be hotter than blackbodies.   
The radius of the outer narrow ring (220$\pm$40\,au), 
that is predominantly responsible for the disk emission at wavelengths $>$50{\micron}, is broadly
consistent with the characteristic disk radius derived from the marginally resolved PACS images.

\subsection{The gas disk around HD\,131835} \label{gasparams}

\subsubsection{Basic gas mass estimates}
\begin{figure} 
\includegraphics[scale=.45,angle=0]{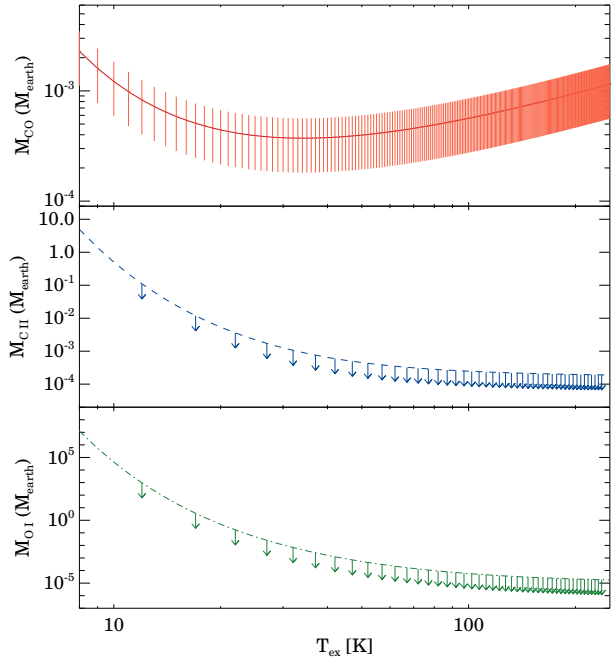}
\caption{ {\sl Gas mass versus excitation temperature for CO, \ion{C}{2}, and \ion{O}{1} in HD\,131835, assuming optically thin 
gas emission.
For \ion{C}{2} and \ion{O}{1} the plotted values are upper limits. }
\label{gasmasses}}
\end{figure}

Observations of a single CO isotopologue ($^{12}$CO in our case) cannot provide information on 
the optical depth of the radiation.
Assuming that the measured CO emission of HD\,131835 is optically thin, the mass of 
CO gas in the disk can be estimated as 
\begin{equation} 
M_{CO} = \frac{4 \pi m d^2}{h \nu_{ul} A_{ul}} \frac{S_{ul}}{x_u},    
\end{equation}
where $m$ is the mass of the CO molecule, $d$ is the distance of the object, $h$ is the Planck constant,
 $\nu_{ul}$ and $A_{ul}$ are the rest frequency and the Einstein coefficient 
 for the given transition between the $u$ upper and $l$ lower levels, $S_{ul}$ is the observed integrated line flux, while 
 $x_u$ is the fractional population of the upper level. Apart from $x_u$ all of the parameters from the right 
 side of the equation are known. In case of local thermodynamic equilibrium (LTE), 
 the level populations are thermalised and governed
by the Boltzmann equation, allowing the computation of $x_u$ if we know 
the gas temperature. In optically thin disks the gas kinetic temperature can significantly differ 
from the temperature of dust grains \citep[see e.g.][]{kamp2001} thus the dust emission does not 
constrain the gas temperature. Since in the 2--1 and 4--3 transitions we have upper limits 
for the CO emission (Fig.~\ref{co32}), from the line ratios we can provide only a poor constrains for the 
excitation temperature, yielding  $T_{\rm ex} > $8\,K.  
Taking $T_{\rm ex} = 20$\,K, the value measured in $\beta$\,Pic \citep{roberge2006} 
we derived a total CO gas mass of 
4.4$\pm$2.2$\times$10$^{-4}$\,M$_\oplus$ from our $J$=3--2 line flux.
In LTE $T_{\rm ex}$ and $T_{\rm kin}$ are equal.
As Fig.~\ref{gasmasses} demonstrates the derived CO mass is not very sensitive to the excitation temperature.
For a range between 8 and 250\,K, $M_{\rm CO}$ varies between 3.7$\times$10$^{-4}$  and 2.3$\times$10$^{-3}$\,M$_\oplus$.
In principle using similar assumptions we can derive upper limits both for \ion{O}{1} and \ion{C}{2} masses, however
for these constituents -- especially for \ion{O}{1} -- the estimates have stronger dependency on the 
excitation temperature (see Fig.~\ref{gasmasses}). 
Additionally, LTE excitation of \ion{O}{1} 63{\micron} line requires a dense medium, 
e.g. considering H$_2$ 
molecules as collisional partners the critical density is typically 
$\sim$5$\times$10$^5$\,cm$^{-3}$ \citep{lequeux2005}. In a lower density 
environment this line becomes subthermal with excitation temperatures far below what would be expected from LTE. 
Since these issues would make the mass estimate of \ion{O}{1} unreliable we focused on \ion{C}{2},
 where the critical densities are lower 
 \citep[e.g. $\sim$3$\times$10$^3$\,cm$^{-3}$, assuming hydrogen molecules as 
 collisional partners,][]{lequeux2005}.  
For $T_{\rm ex} > $8\,K, we obtained 
$M_{\rm C\,II} = 1.4$\,M$_\oplus$ as an upper mass limit 
for \ion{C}{2}.

\subsubsection{Simple gas disk model}
It is important to note, however, that our basic assumptions concerning the optically thin gas emission 
and the LTE could be questioned. For instance, in the disk around HD\,21997 not only the $^{12}$CO but even the 
$^{13}$CO lines 
turned out to be optically thick \citep{kospal2013},  
while in debris disks with low amount of gas,
radiative excitation can dominate over collisional processes
leading to subthermal non-LTE level populations \citep{matra2015}. 
To partly consider these caveats and to further constrain the gas disk fundamental 
properties the measured line profile was modeled with a simple disk geometry using 
the LIME radiation transfer code \citep{brinch2010}. 


We adopted a prescription for the structure of the disk which is frequently used
to model passively irradiated protoplanetary disks. The radial surface density distribution 
was taken to be a powerlaw. We assumed that the disk is vertically isothermal and in 
hydrostatic equilibrium, resulting in a Gaussian vertical density
distribution, with radially dependent pressure scale height. Assuming Keplerian rotation around 
a 1.77\,M$_\odot$ star
and a radial temperature profile of r$^{-0.5}$, the ratio of the pressure scale 
height to the radius has a profile of r$^{-0.25}$. Using these approximations the disk 
can be described as:
\begin{equation}
n_{\rm CO}(r,z)= \frac{{\Sigma}_{\rm CO}}{\sqrt{2\pi}H_{\rm p}} \exp{\left[-\frac{1}{2}\left(\frac{z}{H_{\rm p}}\right)^2\right]},
\end{equation}
\begin{equation}
\frac{H_{\rm p}}{r} = 0.1 \left(\frac{r}{R_{\rm out}}\right)^{-0.25},
\end{equation}
\begin{equation}
{\Sigma}_{\rm CO}(r) = {\Sigma}_{\rm CO,out}\left(\frac{r}{R_{\rm out}}\right)^{-1},
\end{equation}

where $R_{\rm out}$ is the outer disk radius, ${\Sigma}_{\rm CO,out}$ is the gas surface density at the outer disk radius in the 
midplane and $H_{\rm p}$ is the scale height. 
Further parameters are the inner radius ($R_{\rm in}$), the inclination ($i$) of the disk, and the gas 
temperature $T_{\rm g}(r)$. In the modelling we always assumed that the gas disk is coplanar 
with the dust disk and therefore its inclination was fixed to 74{\degr} (Sect.~\ref{dustprops}).
The origin of gas in this system is not clarified yet, it could either be primordial, or secondary. 
In the modelling therefore we used two different scenarios corresponding 
to these alternative hypotheses.

\subsubsection{Primordial gas disk model}
In our first model we assumed a primordial gas disk 
where the CO gas is accompanied with so much molecular hydrogen that the whole disk is in LTE, i.e. the density of H$_2$ exceeds the critical density 
everywhere and the rotational excitation is dominated by collisions.  
Concerning the radial distribution 
first we examined whether a very narrow gas ring co-located with 
 the inner or the outer narrow dust rings proposed by \citet{hung2015} can reproduce 
the observed CO (3--2) spectrum. We found that none of these solutions are feasible because 
the obtained model spectra are significantly narrower than the measured one. 
Then we introduced a more extended gas disk model whose inner radius was fixed to 35\,au  
\citep[corresponding to the inner radius of the proposed continuous dust disk component in the model of][]{hung2015} 
while the outer radius was varied between 80 and 310\,au. The radial distribution of gas temperature was described as 
$T_{\rm g}(r) = T_{\rm g,in} r^{-0.5},$ the temperature at the disk's inner edge, $T_{\rm g,in}$, was also a free parameter 
and varied between 10 and 100\,K. 
We found that only models with $R_{\rm out}$ between 80 and 150\,au provide spectral profiles 
consistent with the measured 3--2 line spectrum. 
 The CO (3--2) line emission turned out to be at least partly optically thick in all models.
The gas temperature cannot be strongly constrained, 
only $T_{\rm g,in} \leq 10\,K$ could be excluded, since with this low temperature the obtained model spectra always found to be too 
faint. 
The lowest CO mass, $M_{\rm CO} = 5.2\times$10$^{-4}$\,M$_\oplus$ was obtained with 
$R_{\rm out} = 120$\,au, ${\Sigma}_{\rm CO,out} = 4.6\times10^{15}$\,cm$^{-2}$ and $T_{\rm g,in} = 56$\,K. 
CO spectra in the relevant transitions belong to this model were plotted in Figure~\ref{co32}. 
By adopting a canonical CO/H$_2$ abundance ratio of 10$^{-4}$ and using this 
minimum mass model we could check our original assumption 
on LTE a posteriori. We found that an overwhelming part (more than 95\%) of 
CO gas is located in disk regions where the density 
of H$_2$ exceeds the critical density of CO J=3--2 transition. 
{Actually, by relaxing our assumption of LTE and repeating the LIME computations 
with the canonical CO/H$_2$ 
abundance ratio for this disk model we obtained CO spectra consistent with the ones derived 
in the LTE assumption. 
These results support that LTE is a reasonable assumption in our models.}

\subsubsection{Secondary gas disk model}
As an alternative scenario, we also examined a disk whose  
gas material is produced from icy grains and planetesimals. 
In this case, the gas would mainly contain ${\rm H_2O}$ and CO, similarly 
to comets in our solar system \citep{mumma2011}.
Since the low dust content of the disk does not provide effective shielding against stellar and interstellar 
UV photons, the released molecules are quickly photodissociated. 
Photodissociation of water molecules 
most commonly produce OH radicals and H atoms, from which
the former are then photodissociated into H and O atoms.
By comparing the number of dissociating photons 
from the stellar photosphere model (Sect.~\ref{stellarprops})
and from the interstellar radiation field \citep[][]{draine1978}, we concluded that 
 the radiation field at the relevant wavelengths of  $<$1900{\AA} is dominated by stellar UV photons everywhere in the disk.
Using photodissociation cross sections from \citet{lee1984}
we found that water molecules are very rapidly photodissociated, their lifetime is lower than 3.4\,days even 
at the outer edge of the disk ($\sim$120\,au). 
Photodissociation of CO molecules requires more energetic UV photons \citep[$<$1118{\AA},][]{visser2009}. Within 
$\sim$45\,au, stellar radiation is the dominant source of these photons, out of this region the  
contribution of the interstellar radiation field is more important.
$^{12}$CO molecules can survive longer, their lifetime is estimated to be $\gtrsim$40\,yr in the 
disk (see Sect.~\ref{gasprops}).
The photodissociation of CO results in O and C atoms. A fraction of carbon atoms then could be ionized by 
the stellar or interstellar UV photons, yielding carbon ions and electrons.

Following the above mentioned considerations, we constructed a disk model containing
well mixed CO and \ion{C}{2} gas. 
From the potential collisional partners we only took into account the electrons. Collisional rate 
coefficients for CO--$e^-$ were derived based on \citet{dickinson1975}, while  
\ion{C}{2}-$e^-$ coefficients were taken from \citet{wilson2002}.
No coefficients are available for CO--O, \ion{C}{2}--O or CO--\ion{C}{1}, \ion{C}{2}--\ion{C}{1} 
collisions.
Based on coefficients for hydrogen atoms from \citet{yang2013} and \citet{barinovs2005}, 
we found that for $n_{\rm H} / n_{\rm e^-} <$ 200 the 
collisions with electrons dominate the excitation of both CO and \ion{C}{2}.
We adopted the same CO gas density distribution as in the primordial model, 
$R_{\rm in}$, $R_{\rm out}$ and $i$ were fixed to 35\,au, 120\,au, and 74{\degr}.
$T_{\rm g,in}$ and the abundance ratio of \ion{C}{2} ions to CO molecules 
($n_{\rm C\,II} / n_{\rm CO}$) were free parameters.
Since H and O are not ionized in regions not subject to extreme UV photons, 
and thereby carbon can be considered as the dominant source of free electrons,  
we adopted $n_{\rm e^-} = n_{\rm C\,II}$ everywhere throughout the disk 
\citep[see also][]{zuckerman2012,matra2015}. 

The stellar parameters and the radial extent of the disk in the 
HD\,131835 system resemble those of $\beta$\,Pic (see Sect.~\ref{gaseousatype}).
Therefore in our modelling we used the parameters of the $\beta$\,Pic system as a 
benchmark. 
Based on recent observations with {\sl Herschel} and ALMA,
the disk around $\beta$\,Pic 
contains 5.5$\times$10$^{-3}$\,M$_\oplus$ of \ion{C}{2}, 7.1$\times$10$^{-3}$\,M$_\oplus$ of \ion{C}{1} 
\citep{cataldi2014} and 
2.85$\times$10$^{-5}$\,M$_\oplus$ of CO gas 
\citep{dent2014}. This corresponds to a \ion{C}{2} to CO abundance ratio of 
$\sim$450, and even if we take into account the 
uncertainites of the mass estimates, the ratio is above 150. 
Therefore we varied $n_{\rm C\,II} / n_{\rm CO} = n_{\rm e} / n_{\rm CO}$ between
1 and 450 in our models. We performed non-LTE radiative transfer modelling.
In the course of modelling first we determined the CO density that reproduces
the observed CO (3--2) line for a certain gas temperature and 
\ion{C}{2} to CO abundance ratio, and then we computed the \ion{C}{2} model flux.
The resulting electron densities in the disk always exceeded the critical electron density for \ion{C}{2}, leading to
LTE level populations.  
For abundance ratios of $n_{\rm C\,II} / n_{\rm CO} > 150$, only models with $T_{\rm g,in} < 20$\,K 
were found to be consistent with the measured upper limit of \ion{C}{2}. 
Because of the low gas temperatures, these models would require the presence of at least 
5$\times$10$^{-3}$\,M$_\oplus$ (${\Sigma}_{\rm CO,out} = 4.4\times10^{16}$\,cm$^{-2}$) of CO gas for the 
reproduction of the observed CO (3--2) line. 
Taking into account the \ion{C}{2} ions and \ion{C}{1} atoms, the total gas mass in these models 
is higher than 0.7\,M$_\oplus$. 
The energy of the upper level for the \ion{C}{2} line is $E_{\rm u} = 91$\,K, thus, for a given
amount of \ion{C}{2} gas, the line emission becomes brighter in higher temperature models.
We found that in models with inner gas temperatures higher than 30\,K, the 
\ion{C}{2} to CO abundance ratio must be $\leq 30$ for producing a \ion{C}{2} line fainter 
than our upper limit. 
{CO gas is excited subthermally in most of the disk 
in these cases}, therefore the reproduction of 
the observed CO (3--2) line requires at least a CO mass of 1$\times$10$^{-3}$\,M$_\oplus$ 
(${\Sigma}_{\rm CO,out} = 8.9\times10^{15}$\,cm$^{-2}$).
In these models the total gas mass of CO, \ion{C}{1}, and \ion{C}{2} together is at the same level, 
or only slightly larger, than in the disk of $\beta$\,Pic.

\subsubsection{Search for accretional signatures}
We investigated whether HD\,131835 shows any signatures of active accretion. 
A possible excess in the Balmer discontinuity can be used to estimate the 
rate of accretion in disks around Herbig Ae stars \citep{muzerolle2004,mendigutia2011}. 
The excess parameter $\Delta D_B$ was calculated as $\Delta D_B = (U - B)_0 - (U - B)_{dered}$ 
where $(U - B)_0$ -- the intrinsic color -- was derived from the Kurucz photospheric model 
of the source using synthetic photometry, while for computation of $(U - B)_{dered}$, the dereddened measured color index,  
we used U, B photometry from the catalogue of \citet{slawson1992}. 
We obtained a $\Delta D_B = -0.045\pm0.047$\,mag, in the computation of uncertainty we took
into account both the measurement errors and the uncertainties in the Kurucz photospheric model 
(the uncertainties in the stellar parameters). 
This calculation clearly shows that there is no
 excess in the Balmer discontinuity.
By calculating a 3-$\sigma$ upper limit of $\sim$0.1 for $\Delta D_B$ and following the outline 
described in \citet{mendigutia2011} we derived an upper limit of $\sim$2$\times$10$^{-8}$\,M$_\odot$yr$^{-1}$ 
for the accretion rate.  
By inspecting the high-resolution optical spectra we found 
that the possible accretional indicator lines of H$_\alpha$, H$_\beta$, H$_\gamma$, and He\,I 5876\AA{} 
are in absorption and are consistent with our fitted spectral model. 
For the H$_\alpha$ line luminosity we derived an upper limit of 2$\times$10$^{-4}$\,L$_\odot$. 
Based on the calibration obtained for HAeBe stars by \citet{mendigutia2011} this upper limit corresponds to
an accretion luminosity of $\log{\frac{L_{acc}}{L_{\odot}}} < -1.76$. Considering that 
$L_{acc} = GM_*\dot{M}_{acc} / R_*$, from the absence of H$_\alpha$ line excess
we obtained an upper limit of 
$\dot{M}_{acc} < 5 \times 10^{-10}$\,M$_\odot$~yr$^{-1}$ 
for the mass accretion.

\subsection{Gas mass upper limits for the other targets}
For the non-detected sources CO mass upper limits were estimated assuming optically thin 
emission and local thermodynamic equilibrium (LTE). The excitation temperature was 
assumed to be 20\,K, the line flux upper limits were taken from Table~\ref{basictable}. 
In those cases where both CO (3--2) and (4--3) line observations were 
available we used the lower transition in the calculations.
We note again that our basic assumptions might not be fulfilled for all 
of the studied systems. With the achieved sensitivity we cannot exclude that some of our targets 
harbor optically thick gas disks. In very tenuous gas disks the excitation could be subthermal even leading 
to a very low excitation temperature of $<$10\,K \citep[e.g.][]{matra2015}. In both of these cases our upper limits would 
underestimate the possible total CO mass in the disk.

\section{Discussion}
\label{discussion}

\subsection{Gaseous debris disks around young A-type stars} \label{gaseousatype}


With the discovery of CO gas in HD\,131835 the number of 
known gaseous debris disks around  A-type stars is increased to seven. 
Tables~\ref{gaseousdisks1} and \ref{gaseousdisks2} present the main 
stellar and disk properties. 
All of these systems are likely younger than 50\,Myr\footnote 
{\citet{rodigas2014} estimated that the age of 
HD\,32297 falls
between 15 and 500\,Myr, but, as they and \citet{donaldson2013} noted, 
the kinematic properties of the star 
and the fractional luminosity of the disk imply an age close to the 
lower limit of this interval.}, thus they represent
the very early phase of debris disk evolution.
HD\,131835 may be one of the youngest objects in this sample
and based on their nearly identical stellar properties, 
 it can be considered as a comparably young sibling 
 of $\beta$\,Pic.

\begin{figure} 
\includegraphics[scale=.45,angle=0]{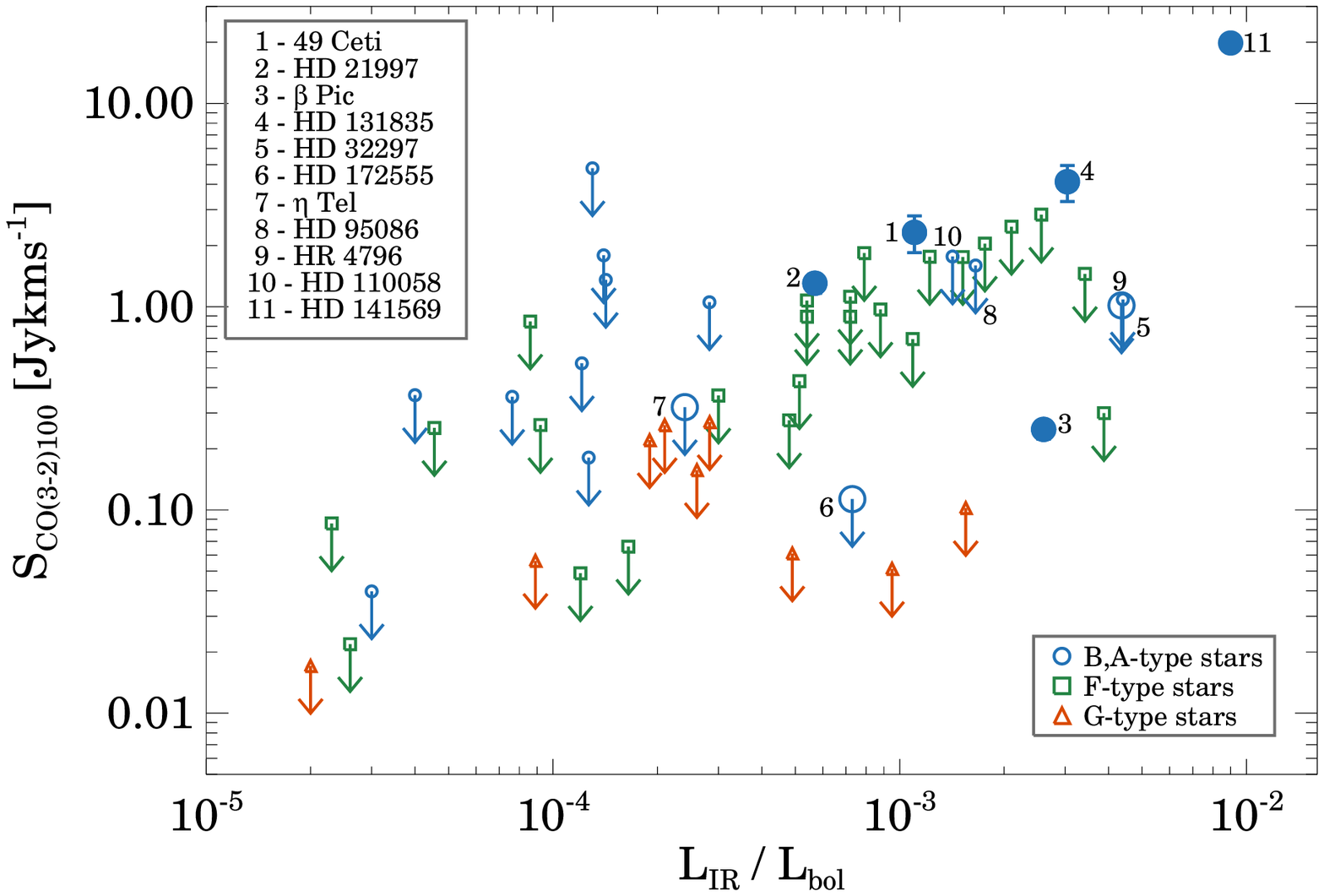}
\caption{ {\sl Integrated CO(3--2) fluxes or 
upper limits 
for debris disks and the transitional disk around HD\,141569
normalized to 100\,pc are plotted against fractional luminosities.
{ 
Fractional luminosities and integrated line fluxes for the annotated 
sources were taken from: 
49\,Ceti -- \citet{moor2015}, \citet{dent2005}; 
HD\,21997 -- \citet{moor2015}, \citet{kospal2013};
$\beta$\,Pic -- \citet{rhee2007}, \citet{dent2014};
HD\,32297 -- \citet{chen2014}, \citet{moor2011b};
HD\,172555 -- \citet{rm2012}, \citet{moor2011b};
HR\,4796 -- \citet{chen2014}, \citet{greaves2000a};
HD\,110058 -- \citet{chen2014}, \citet{moor2011b};
HD\,141569 -- \citet{meeus2012}, \citet{dent2005};
HD\,131835, $\eta$\,Tel, HD\,95086 -- Table~\ref{basictable}. 
For the rest of the objects the data are from 
\citet{ballering2013,chen2005,chen2014,dent2005,fajardo1998,hales2014,
moor2011a,moor2011b,moor2015,najita2005,patel2014,pascucci2006,rhee2007}. }
Disks with known gas component are denoted by larger symbols.
}
\label{sco100}
}
\end{figure}

Figure~\ref{sco100} shows the integrated CO(3--2) fluxes
and upper limits for these seven sources, 
normalized to 100\,pc, and plotted against fractional luminosities.
Partly based on our two surveys, additional debris disks around 
main-sequence stars located within 125\,pc, 
as well as HD\,141569, a nearby transitional disk that may represent a very 
final phase of protoplanetary disk evolution \citep{wyatt2015}, 
are also displayed. Although there is no clear trend 
with the fractional luminosity, 
apart from $\eta$\,Tel all known gaseous debris disks have a fractional 
luminosity $>$5$\times$10$^{-4}$. Based on debris disk catalogues of 
\citet{moor2006}, \citet{rhee2007}, and \citet{chen2014} we identified eleven A-type stars 
within 125\,pc that harbor disks with $L_{\rm IR}$ / $L_{\rm bol}$ above 5$\times$10$^{-4}$. 
 All of these systems are younger 
than 50\,Myr and
nearly all of them have already been observed in CO rotational transitions 
in different surveys
(the two exceptions are HD\,98363 and HD\,143675). Among the observed nine disks, four 
harbor CO gas (49\,Ceti, HD 21997, $\beta$\,Pic and HD\,131835), 
while in the case of HD\,32297 and HD\,172555 
atomic gas tracers were detected. The gas component of HD\,95086, HR\,4796, and HD\,110058 have 
remained undetected so far.
It suggests that the most massive
debris disks around young A-type stars quite commonly possess 
detectable amount of gas. 
It is important to note that among these 
disks only HD\,21997 and $\beta$\,Pic were observed with ALMA and 
the sensitivity of most of the other observations did not allow the 
detection of even a $\beta$\,Pic-like gas disk.  
From the four CO bearing debris disks HD\,131835 is the brightest, its 
normalized line flux is $\sim$17 times higher than that of the disk around 
$\beta$\,Pic, but falls below the normalized line flux of HD\,141569. 

Within 125\,pc there are many F-type stars with age of $<$50\,Myr that 
possess debris disks with a fractional luminosity of $>$5$\times$10$^{-4}$ (Fig.~\ref{sco100}), 
i.e. having similar properties than that of the abovementioned A-type sample. 
We observed 15 such systems in the framework of our two surveys but 
none of them were detected, thus based on the current data sets 
the presence of CO molecules may be 
the characteristic of debris disks around young A-type stars. 
{\citet{rigliaco2015} detected H\,I lines in mid-infrared spectrum of
eight young debris disks around F-K type stars.
The presence of these lines can be explained either by 
low rate gas accretion onto these stars or chromospheric 
activity. In the case of former scenario these findings 
would indicate that some disks around F-K type stars can also retain 
primordial gas for their debris phase.}

\subsection{Dust properties of the HD\,131835 disk} \label{diskclass}
Based on its infrared and submillimeter continuum data the 
dust disk properties of HD\,131835 are similar to 
those of the other young debris disks. Although the disk 
has a high total fractional luminosity of $\sim$3$\times$10$^{-3}$ 
this value is still less than the usually invoked threshold value of 0.01, 
that divides protoplanetary and debris disks. 
Recently, \citet{wyatt2015} introduced a new classification scheme
to differentiate between Herbig Ae disks and debris disks with A-type host stars.
This method is based on the 
ratios of the observed to the stellar photospheric fluxes at 12 and 70{\micron} ($R_{12}$ and $R_{70}$).
For debris disks $R_{12}$ and $R_{70}$ have to be lower than 3 and 2000, respectively. 
Based on color-corrected WISE 12 and PACS 70{\micron} data and predicted photospheric fluxes
we derived $R_{12} = 1.8$ and $R_{70} = 955$ for HD\,131835, i.e. 
it can be classified as a debris disk. 
The majority of Herbig Ae systems show PAH emission. The absence of PAH features in the IRS spectrum 
of HD\,131835 is also in accordance with its classification as a debris disk.  
These arguments suggest that the observed excess emission is due to dust grains that are probably
second generation, 
 produced via erosion of larger unseen bodies.

\subsection{The gas disk of HD\,131835} \label{gasprops}
According to the current paradigm the gas content of debris disks may also 
be second generation, derived from previously formed planetesimals. 
However, in young disks we cannot exclude the possibility 
that the evolution of dust and gas were not parallel and the observed 
gas is predominantly composed of residual primordial material
(of course we never exclude 
that a part of the gas is produced from already emerged icy bodies). 
Indeed, though most known gaseous 
debris disks are proposed to be rather secondary, one of the oldest among them, 
HD\,21997 may likely harbor a hybrid disk with a significant amount of 
primordial gas \citep{kospal2013}. 
In the following, we investigate the origin of gas in the HD\,131835 system.
In this analysis we considered the 
$\beta$\,Pic and HD\,21997 systems as references:
the former may be the best established representantive  
of young debris disks with secondary gas \citep{fernandez2006,dent2014}, 
while the latter
 is the sole known example for a hybrid disk. Furthermore, 
 both systems have already been observed with ALMA \citep{dent2014,kospal2013} providing very detailed 
 information on the spatial distribution of CO gas and dust,  
 and their host stars -- particularly $\beta$\,Pic -- 
 is quite similar to HD\,131835.

\begin{figure} 
\includegraphics[scale=.45,angle=0]{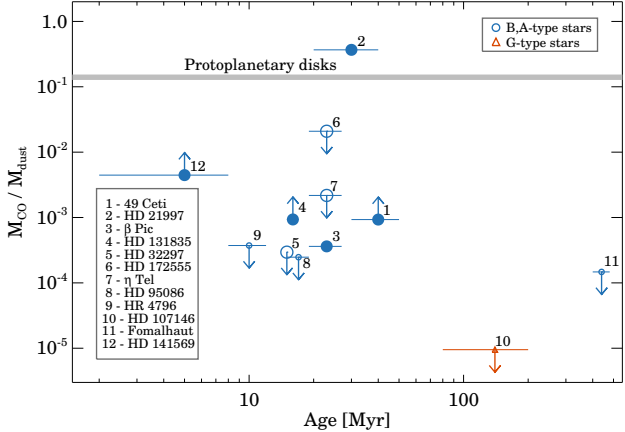}
\caption{ {\sl CO to dust mass ratio as a function of age for debris disks and for HD\,141569. 
Debris disks with known gas component are marked by larger symbols, those where CO gas 
was detected are shown by filled symbols. Their CO/dust mass data and ages were taken from 
{Table~\ref{gaseousdisks1} and \ref{gaseousdisks2}}. 
HD\,32297 was placed at its lower age limit proposed by \citet{rodigas2014}.
For Fomalhaut and HD\,107146 we used data from the literature 
\citep[][and references therein]{matra2015,ricci2015}, in the latter case the CO upper mass estimate 
derived from the secondary disk scenario were taken into account.   
In the case of HD\,141569 and HR\,4796 the CO mass and 
upper limit were 
computed by adopting an excitation temperature of 20\,K and using
the CO (3--2) line fluxes from \citet{dent2005,hales2014}. Their dust masses were computed 
from their submillimeter fluxes \citep{sheret2004} using the standard way. 
{Age estimate of HD\,141569 was taken from \citet{weinberger2000}.
As a member of the TW\,Hya association, for HR\,4796 we applied the age of the group \citep{torres2008,gagne2014}. }
The data of HD\,95086 are from Table~\ref{basictable} and \citet{moor2015}.
The typical $M_{\rm CO} / M_{\rm dust}$ ratio for protoplanetary disks (horizontal line) 
was calculated by assuming a total gas to dust mass ratio of 100 and taking a canonical 
CO to H$_2$ abundance ratio of 10$^{-4}$. 
}
\label{mcomdust}
}
\end{figure}

In the absence of significant H$_2$ and dust content 
in a debris disk with secondary gas, only self-shielding can protect CO molecules 
againt the UV photons of stellar and interstellar radiation field.
To judge the efficiency of self-shielding we took  
our secondary gas disk model with a CO mass of 1.0$\times$10$^{-3}$\,M$_\oplus$ (i.e. the model 
with the minimum CO mass)
and computed the shielding factors using the photodissociation 
model of \citet{visser2009}. The stellar UV flux found to be dominant at $<$45\,au, were determined from the fitted 
ATLAS model (Sect.~\ref{stellarprops}). The contribution of interstellar radiation field 
  were computed from \citet{draine1978}. 
{Our calculations yielded a CO lifetime of 40\,yr 
at the directly illuminated inner edge of the disk and of 500\,yr 
in the most efficiently self-shielded mid-plane regions.} 
Assuming that the gas is continuously replenished and taking the CO mass of 1.0$\times$10$^{-3}$\,M$_\oplus$
we obtained a CO production rate of 
$\gtrsim$ 
2$\times$10$^{-6}$\,M$_\oplus$yr$^{-1}$ 
or 1.2$\times$10$^{19}$\,kg~yr$^{-1}$. 
This is at least eight times higher than the gas production rate obtained for the $\beta$\,Pic disk from its 
ALMA observation \citep{dent2014}. Assuming a CO mass abundance of 10\% in 
planetesimals, it would require destruction at least of 2$\times$10$^{-5}$\,M$_\oplus$yr$^{-1}$ of icy bodies.  
The CO gas can be released from icy grains/planetesimals in different processes.
The temperature significantly exceeds the value needed for the sublimation of pure CO ice ($\sim$20\,K) 
everywhere in the disk. In such environment CO ice is thought to be mainly present in deeper layers of icy 
planetesimals and as admixture in the amorphous water ice on the surface of grains and larger bodies.
Thus photodesorption or, in warmer regions ($>$ 110\,K), sublimation of surface water ices can produce CO gas 
as well. CO entrapped in water ice matrix can also be released via collisions between grains and planetesimals, moreover 
fragmentation of larger bodies can lead to the excavation of CO ices persisted in deeper layers.
Figure~\ref{mcomdust} shows the ratio of CO mass to the dust mass as a function of age for 
debris disks with known gaseous component and for some selected disks (additional 
debris disks around A-type stars with $f_{\rm d} > 5\times10^{-4}$; Fomalhaut and HD\,107146, two debris systems recently 
observed by ALMA; and HD\,141569). Based on ALMA observations of \citet{dent2014} 
the $\frac{M_{\rm CO}}{M_{\rm dust}}$ ratio
measured for $\beta$\,Pic disk is
 3.6$\times10^{-4}$ \citep[note that they used an assumption of LTE, and if it is not fulfilled 
the CO mass could be higher, see also in][]{matra2015}. 
 The spatially resolved images also revealed a clump at $\sim$85\,au 
 from the star, in which an enhanced 
 CO-to-dust mass ratio was measured.
Sensitive ALMA observations are available only for a few debris disks. Upper limits 
for $\frac{M_{\rm CO}}{M_{\rm dust}}$ at Fomalhaut \citep{matra2015} and at 
HD\,107146 \citep{ricci2015}, however, hint that the ratio measured at $\beta$\,Pic 
is already quite high. 
Interestingly the other three debris disks with CO gas show even higher $\frac{M_{\rm CO}}{M_{\rm dust}}$ ratios. 
For HD\,21997 we measured a $\sim$1000 times higher $\frac{M_{\rm CO}}{M_{\rm dust}}$ ratio 
(note that here we used 
the optically thin $^{12}$C$^{18}$O line to estimate the CO mass) than for the $\beta$~Pic. 
This object clearly differs from the others and likely harbors a significant fraction of 
primordial gas. 
The lower limits of CO-to-dust mass ratios for HD\,131835 (in calculation of its $\frac{M_{\rm CO}}{M_{\rm dust}}$ we 
used the minimum CO mass of 5.2$\times$10$^{-4}$\,M$_\oplus$ that we obtained in our models)
 and 49\,Ceti are $\gtrsim$2.5$\times$ higher 
than that of $\beta$\,Pic.
In a secondary disk the $\frac{M_{\rm CO}}{M_{\rm dust}}$ ratio depends both on the dust and gas production rates and 
on the removal time of the two elements. 
The observed higher CO-to-dust mass ratios thus can be explained e.g. by 
a more effective self-shielding due to larger CO density which 
would result in higher CO lifetime or by higher fraction of volatile material in icy 
planetesimals/grains. 

\begin{figure} 
\includegraphics[scale=.45,angle=0]{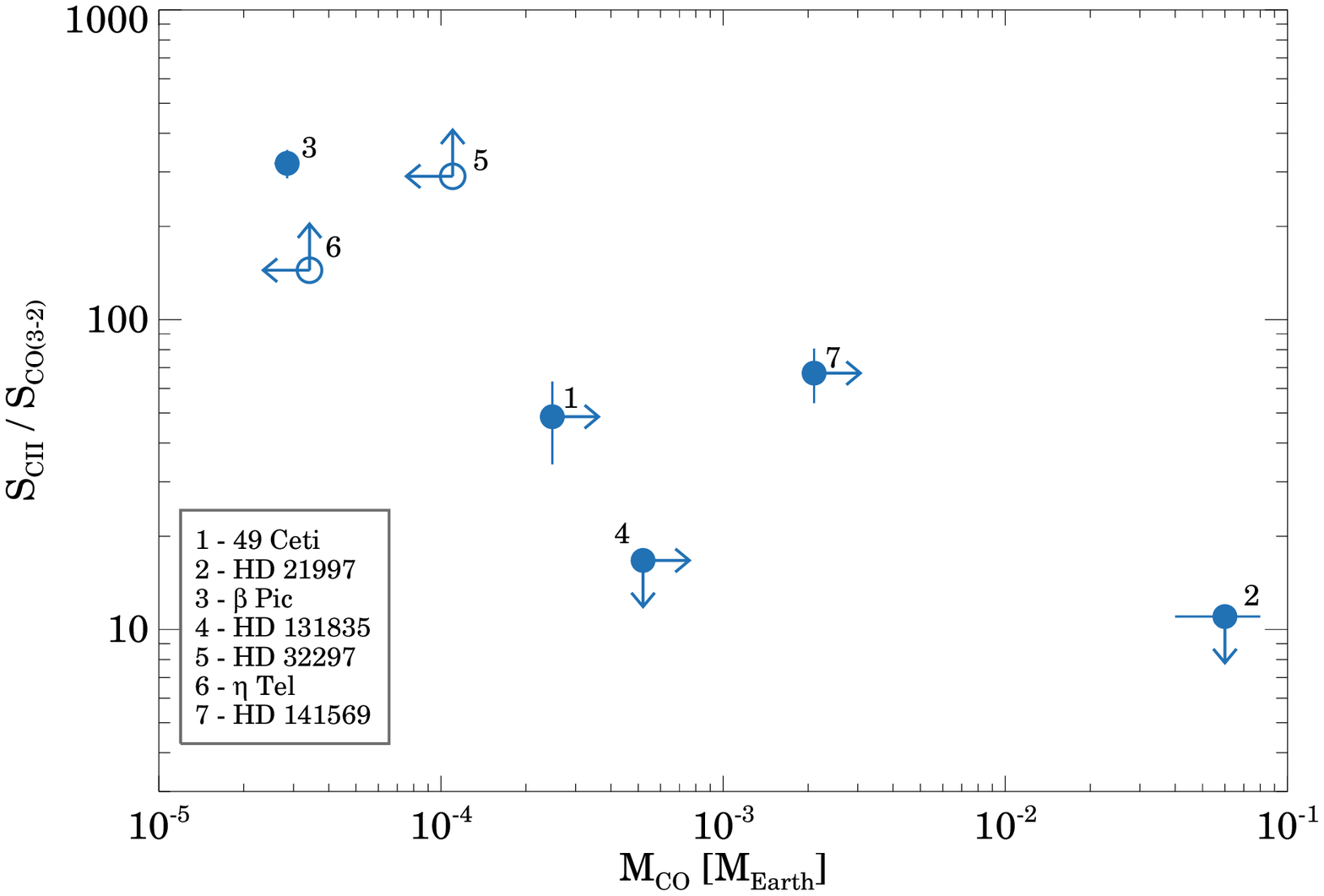}
\caption{ {\sl S$_{C\,II}$/S$_{CO(3-2)}$ ratio as a function of total CO mass for gaseous 
debris disks (apart from HD\,172555 where neither of CO, nor \ion{C}{2} were detected) and 
HD\,141569. \ion{C}{2} line flux data for HD\,131835 and HD\,21997 were taken from Sect~\ref{herschelobs} and  
from Juh\'asz et al. (in prep.) For the other targets S$_{C\,II}$ values were from  
49\,Ceti -- \citet{roberge2013}; $\beta$\,Pic -- \citet{cataldi2014}; HD\,32297 --  \citet{donaldson2013}; 
$\eta$\,Tel -- \citet{rm2014}; HD\,141569 --  \citet{meeus2012}. References for the S$_{CO(3-2)}$ data 
are summarized in Fig.~\ref{sco100} caption.
}
\label{cii}
}
\end{figure}

Stellar and interstellar UV photons can permeate debris disks without any hindrance 
and ionize carbon atoms formed via photodissociation of CO molecules. 
The fraction of neutral and ionized carbon depends on the strength of the local radiation field. 
Due to their longer lifetime in such environment, the amount of \ion{C}{1} and \ion{C}{2} 
can significantly exceed that of CO molecules, e.g. in the disk of $\beta$\,Pic 
there is approximately 200 times more \ion{C}{2} gas than CO gas in terms of mass \citep{cataldi2014,dent2014}.
Figure~\ref{cii} displays the $S_{\rm C\,II}$/$S_{\rm CO(3-2)}$ line flux ratios 
for debris disks where C\,II and/or CO gas has been detected. 
The ratio of $\beta$\,Pic is about 30 times
higher than the upper limit we obtained for HD\,131835. 
This may reflect a real difference in the mass ratio, or -- as our modelling showed (Sect.~\ref{gasparams}) -- 
may be a consequence of the low 
gas temperature in HD 131835. 
In the latter models the total gas mass and the gas-to-dust mass ratio would be rather high. 

In the secondary gas scenario, gas and dust
are released at the same location from planetesimals, thus
the two components are expected to be co-located, just as we see at $\beta$\,Pic.
Based on our current data and modelling (Sect.~\ref{gasparams}) we cannot determine the accurate location 
of gas, we just can say that the results do no exclude that the two components are co-located.
It is worth noting, however that our gas disk model coincides
with that part of the dust disk model of \citet{hung2015}, which is composed by 
 very hot grains. These grains give only a low fraction of the dust disk mass and it is questionable 
  whether they are accompanied with large amount planetesimals that could be the  
  source of observed CO gas.

All in all, in the secondary
scenario, HD\,131835 could be considered as of comparable age and -- in terms of its disk -- 
more massive analogue of the $\beta$\,Pic system. 
However, none of our current data exclude that HD\,131835 in fact possess a 
hybrid disk. Our modelling in Sect.~\ref{gasparams} shows that the CO (3--2) emission may be optically thick, thus 
we cannot exclude that the CO mass is significantly higher reaching even the same level as in the case of 
HD\,21997. The low $S_{\rm C\,II}$/$S_{\rm CO(3-2)}$ flux ratio can be explained in the secondary scenario, but  
as Figure~\ref{cii} shows, among the currently known gaseous debris disks 
only HD\,21997, the sole hybrid disk, exhibits similarly low ratio. 

For a better clarification of the origin of gas additional 
observations are necessary. By measuring the emission 
of less abundant $^{13}$CO and C$^{18}$O isotopologues from the 
disk a more reliable CO gas mass estimate could be 
obtained.  
Moreover, by mapping and comparing the spatial distribution of gas and dust
allow us to investigate whether the two components 
are co-located, which is a pre-requisite in a secondary 
gas scenario. These observations requires better 
sensitivity and spatial resolution than that of a 
single dish radio telescope but could definitely be 
carried out using the Atacama Large Millimeter/submillimeter 
Array.

\section{Summary} \label{summary}
By searching for CO gas in 20 debris disks using the APEX and IRAM\,30m radiotelescopes 
we identified a new gas rich system around the $\sim$16\,Myr old UCL member, HD\,131835, where the 3--2 transition of CO 
was successfully detected. 
Based on spectroscopic and photometric data the stellar properties of HD\,131835 resembles well 
those of $\beta$\,Pic. By observing HD\,131835 with the {\sl Herschel Space Observatory} 
we found that the disk is spatially resolved both at 70 and 100{\micron} and the disk characteristic radius 
is $\sim$170\,au. Thanks to our observations at 250, 350, and 500{\micron} the submillimeter SED of the source 
is now significantly better characterized. 
Based on its infrared and submillimeter continuum data the 
dust disk properties of HD\,131835 are similar to 
those of the most massive young debris disks.
{With the detection of gas in HD\,131835 the number of 
of known debris disks with CO component is increased to four (49\,Ceti, HD 21997, $\beta$\,Pic and HD\,131835).} 
All of these disks
encircle young ($\leq$40\,Myr) A-type stars. 
Within 125\,pc we know 11 A-type stars whose debris disks' fractional luminosity exceed $5\times$10$^{-4}$. 
Among these disks, 9 have already observed in CO rotational transitions. Four of them
harbor CO gas,  
while in two other objects 
atomic gas tracers were detected. This detection rate of 4/9 (or 6/9 if atomic gas detection was also taken into account)
 suggests that the most massive
debris disks around young A-type stars commonly possess 
detectable amount of gas. Based on our current data we cannot draw a secure conclusion on 
the origin of gas in HD\,131835. If the gas is secondary
then HD\,131835 could be considered as a comparably young and -- in terms of its disk -- 
more massive analogue of the $\beta$\,Pic system. However, we cannot exclude that this system
-- similarly to HD\,21997 -- possess a hybrid disk, where the gas material is predominantly 
 primordial, while the dust grains are mostly derived from planetesimals.

\acknowledgments
We thank the anonymous referee for useful comments that
helped us to improve the manuscript.
This work was supported by the Momentum grant of the MTA CSFK Lend\"ulet Disk Research Group, 
the PECS-98073 program of the European Space Agency (ESA) 
and the Hungarian Research Fund OTKA grants K101393 and K104607.
A.M. acknowledges support from the Bolyai Research Fellowship of the Hungarian 
Academy of Sciences. C.G. acknowledges support under the 
NASA Origins of Solar System Program on NNG13PB64P. 
A.J. acknowledges the support of the DISCSIM project, grant
agreement 341137 funded by the European Research Council under
ERC-2013-ADG.
This publication makes use of data products from the Wide-field
Infrared Survey Explorer, which is a joint project of the University
of California, Los Angeles, and the Jet Propulsion
Laboratory/California Institute of Technology, funded by the National
Aeronautics and Space Administration. The publication also makes use
of data products from the Two Micron All Sky Survey, which is a joint
project of the University of Massachusetts and the Infrared
Processing and Analysis Center/California Institute of Technology,
funded by the National Aeronautics and Space Administration and the
National Science Foundation.

{\it Facilities:} \facility{APEX}, \facility{IRAM:30m}, \facility{Herschel}, \facility{Spitzer}.



\begin{thebibliography}{}
\expandafter\ifx\csname natexlab\endcsname\relax\def\natexlab#1{#1}\fi

\bibitem[{{Alexander} {et~al.}(2014){Alexander}, {Pascucci}, {Andrews},
  {Armitage}, \& {Cieza}}]{alexander2014}
{Alexander}, R., {Pascucci}, I., {Andrews}, S., {Armitage}, P., \& {Cieza}, L.
  2014, Protostars and Planets VI, 475

\bibitem[{{Backman} \& {Paresce}(1993)}]{backman1993}
{Backman}, D.~E., \& {Paresce}, F. 1993, in Protostars and Planets III, ed.
  E.~H. {Levy} \& J.~I. {Lunine}, 1253--1304

\bibitem[{{Ballering} {et~al.}(2013){Ballering}, {Rieke}, {Su}, \&
  {Montiel}}]{ballering2013}
{Ballering}, N.~P., {Rieke}, G.~H., {Su}, K.~Y.~L., \& {Montiel}, E. 2013,
  \apj, 775, 55

\bibitem[{{Balog} {et~al.}(2014){Balog}, {M{\"u}ller}, {Nielbock}, {Altieri},
  {Klaas}, {Blommaert}, {Linz}, {Lutz}, {Mo{\'o}r}, {Billot}, {Sauvage}, \&
  {Okumura}}]{balog2013}
{Balog}, Z., {M{\"u}ller}, T., {Nielbock}, M., {et~al.} 2014, Experimental
  Astronomy, 37, 129

\bibitem[{{Barinovs} {et~al.}(2005){Barinovs}, {van Hemert}, {Krems}, \&
  {Dalgarno}}]{barinovs2005}
{Barinovs}, {\u G}., {van Hemert}, M.~C., {Krems}, R., \& {Dalgarno}, A. 2005,
  \apj, 620, 537

\bibitem[{{Bendo} {et~al.}(2013){Bendo}, {Griffin}, {Bock}, {Conversi},
  {Dowell}, {Lim}, {Lu}, {North}, {Papageorgiou}, {Pearson}, {Pohlen},
  {Polehampton}, {Schulz}, {Shupe}, {Sibthorpe}, {Spencer}, {Swinyard},
  {Valtchanov}, \& {Xu}}]{bendo2013}
{Bendo}, G.~J., {Griffin}, M.~J., {Bock}, J.~J., {et~al.} 2013, \mnras, 433,
  3062

\bibitem[{{Beust} {et~al.}(1994){Beust}, {Vidal-Madjar}, {Ferlet}, \&
  {Lagrange-Henri}}]{beust1994}
{Beust}, H., {Vidal-Madjar}, A., {Ferlet}, R., \& {Lagrange-Henri}, A.~M. 1994,
  \apss, 212, 147

\bibitem[{{Booth} {et~al.}(2013){Booth}, {Kennedy}, {Sibthorpe}, {Matthews},
  {Wyatt}, {Duch{\^e}ne}, {Kavelaars}, {Rodriguez}, {Greaves}, {Koning},
  {Vican}, {Rieke}, {Su}, {Moro-Mart{\'{\i}}n}, \& {Kalas}}]{booth2013}
{Booth}, M., {Kennedy}, G., {Sibthorpe}, B., {et~al.} 2013, \mnras, 428, 1263

\bibitem[{{Brinch} \& {Hogerheijde}(2010)}]{brinch2010}
{Brinch}, C., \& {Hogerheijde}, M.~R. 2010, \aap, 523, A25

\bibitem[{{Carter} {et~al.}(2012){Carter}, {Lazareff}, {Maier}, {Chenu},
  {Fontana}, {Bortolotti}, {Boucher}, {Navarrini}, {Blanchet}, {Greve}, {John},
  {Kramer}, {Morel}, {Navarro}, {Pe{\~n}alver}, {Schuster}, \&
  {Thum}}]{carter2012}
{Carter}, M., {Lazareff}, B., {Maier}, D., {et~al.} 2012, \aap, 538, A89

\bibitem[{{Castelli} \& {Kurucz}(2003)}]{castelli2003}
{Castelli}, F., \& {Kurucz}, R.~L. 2003, in IAU Symposium, Vol. 210, Modelling
  of Stellar Atmospheres, ed. N.~{Piskunov}, W.~W. {Weiss}, \& D.~F. {Gray},
  20P

\bibitem[{{Cataldi} {et~al.}(2014){Cataldi}, {Brandeker}, {Olofsson},
  {Larsson}, {Liseau}, {Blommaert}, {Fridlund}, {Ivison}, {Pantin},
  {Sibthorpe}, {Vandenbussche}, \& {Wu}}]{cataldi2014}
{Cataldi}, G., {Brandeker}, A., {Olofsson}, G., {et~al.} 2014, \aap, 563, A66

\bibitem[{{Chen} {et~al.}(2011){Chen}, {Mamajek}, {Bitner}, {Pecaut}, {Su}, \&
  {Weinberger}}]{chen2011}
{Chen}, C.~H., {Mamajek}, E.~E., {Bitner}, M.~A., {et~al.} 2011, \apj, 738, 122

\bibitem[{{Chen} {et~al.}(2014){Chen}, {Mittal}, {Kuchner}, {Forrest}, {Lisse},
  {Manoj}, {Sargent}, \& {Watson}}]{chen2014}
{Chen}, C.~H., {Mittal}, T., {Kuchner}, M., {et~al.} 2014, \apjs, 211, 25

\bibitem[{{Chen} {et~al.}(2012){Chen}, {Pecaut}, {Mamajek}, {Su}, \&
  {Bitner}}]{chen2012}
{Chen}, C.~H., {Pecaut}, M., {Mamajek}, E.~E., {Su}, K.~Y.~L., \& {Bitner}, M.
  2012, \apj, 756, 133

\bibitem[{{Chen} {et~al.}(2005){Chen}, {Patten}, {Werner}, {Dowell},
  {Stapelfeldt}, {Song}, {Stauffer}, {Blaylock}, {Gordon}, \&
  {Krause}}]{chen2005}
{Chen}, C.~H., {Patten}, B.~M., {Werner}, M.~W., {et~al.} 2005, \apj, 634, 1372

\bibitem[{{Crawford}(1975)}]{crawford1975}
{Crawford}, D.~L. 1975, \aj, 80, 955

\bibitem[{{Crawford}(1979)}]{crawford1979}
{Crawford}, D.~L. 1979, \aj, 84, 1858

\bibitem[{{Crifo} {et~al.}(1997){Crifo}, {Vidal-Madjar}, {Lallement}, {Ferlet},
  \& {Gerbaldi}}]{crifo1997}
{Crifo}, F., {Vidal-Madjar}, A., {Lallement}, R., {Ferlet}, R., \& {Gerbaldi},
  M. 1997, \aap, 320, L29

\bibitem[{{Cutri} {et~al.}(2003){Cutri}, {Skrutskie}, {van Dyk}, {Beichman},
  {Carpenter}, {Chester}, {Cambresy}, {Evans}, {Fowler}, {Gizis}, {Howard},
  {Huchra}, {Jarrett}, {Kopan}, {Kirkpatrick}, {Light}, {Marsh}, {McCallon},
  {Schneider}, {Stiening}, {Sykes}, {Weinberg}, {Wheaton}, {Wheelock}, \&
  {Zacarias}}]{cutri2003}
{Cutri}, R.~M., {Skrutskie}, M.~F., {van Dyk}, S., {et~al.} 2003, VizieR Online
  Data Catalog, 2246, 0

\bibitem[{{Czechowski} \& {Mann}(2007)}]{cm2007}
{Czechowski}, A., \& {Mann}, I. 2007, \apj, 660, 1541

\bibitem[{{de Zeeuw} {et~al.}(1999){de Zeeuw}, {Hoogerwerf}, {de Bruijne},
  {Brown}, \& {Blaauw}}]{dezeeuw1999}
{de Zeeuw}, P.~T., {Hoogerwerf}, R., {de Bruijne}, J.~H.~J., {Brown}, A.~G.~A.,
  \& {Blaauw}, A. 1999, \aj, 117, 354

\bibitem[{{Dent} {et~al.}(2005){Dent}, {Greaves}, \& {Coulson}}]{dent2005}
{Dent}, W.~R.~F., {Greaves}, J.~S., \& {Coulson}, I.~M. 2005, \mnras, 359, 663

\bibitem[{{Dent} {et~al.}(2013){Dent}, {Thi}, {Kamp}, {Williams}, {Menard},
  {Andrews}, {Ardila}, {Aresu}, {Augereau}, {Barrado y Navascues}, {Brittain},
  {Carmona}, {Ciardi}, {Danchi}, {Donaldson}, {Duchene}, {Eiroa}, {Fedele},
  {Grady}, {de Gregorio-Molsalvo}, {Howard}, {Hu{\'e}lamo}, {Krivov},
  {Lebreton}, {Liseau}, {Martin-Zaidi}, {Mathews}, {Meeus},
  {Mendigut{\'{\i}}a}, {Montesinos}, {Morales-Calderon}, {Mora}, {Nomura},
  {Pantin}, {Pascucci}, {Phillips}, {Pinte}, {Podio}, {Ramsay}, {Riaz},
  {Riviere-Marichalar}, {Roberge}, {Sandell}, {Solano}, {Tilling}, {Torrelles},
  {Vandenbusche}, {Vicente}, {White}, \& {Woitke}}]{dent2013}
{Dent}, W.~R.~F., {Thi}, W.~F., {Kamp}, I., {et~al.} 2013, \pasp, 125, 477

\bibitem[{{Dent} {et~al.}(2014){Dent}, {Wyatt}, {Roberge}, {Augereau},
  {Casassus}, {Corder}, {Greaves}, {de Gregorio-Monsalvo}, {Hales}, {Jackson},
  {Hughes}, {Lagrange}, {Matthews}, \& {Wilner}}]{dent2014}
{Dent}, W.~R.~F., {Wyatt}, M.~C., {Roberge}, A., {et~al.} 2014, Science, 343,
  1490

\bibitem[Dickinson \& Richards(1975)]{dickinson1975} Dickinson, A.~S., \& Richards, D.
\ 1975, Journal of Physics B Atomic Molecular Physics, 8, 2846


\bibitem[{{Donaldson} {et~al.}(2013){Donaldson}, {Lebreton}, {Roberge},
  {Augereau}, \& {Krivov}}]{donaldson2013}
{Donaldson}, J.~K., {Lebreton}, J., {Roberge}, A., {Augereau}, J.-C., \&
  {Krivov}, A.~V. 2013, \apj, 772, 17

\bibitem[{{Draine}(1978)}]{draine1978}
{Draine}, B.~T. 1978, \apjs, 36, 595

\bibitem[{{Fajardo-Acosta} {et~al.}(1998){Fajardo-Acosta}, {Stencel}, \&
  {Backman}}]{fajardo1998}
{Fajardo-Acosta}, S.~B., {Stencel}, R.~E., \& {Backman}, D.~E. 1998, \apjl,
  503, L193

\bibitem[{{Ferlet} {et~al.}(1987){Ferlet}, {Vidal-Madjar}, \&
  {Hobbs}}]{ferlet1987}
{Ferlet}, R., {Vidal-Madjar}, A., \& {Hobbs}, L.~M. 1987, \aap, 185, 267

\bibitem[{{Fern{\'a}ndez} {et~al.}(2006){Fern{\'a}ndez}, {Brandeker}, \&
  {Wu}}]{fernandez2006}
{Fern{\'a}ndez}, R., {Brandeker}, A., \& {Wu}, Y. 2006, \apj, 643, 509

\bibitem[{{France} {et~al.}(2007){France}, {Roberge}, {Lupu}, {Redfield}, \&
  {Feldman}}]{france2007}
{France}, K., {Roberge}, A., {Lupu}, R.~E., {Redfield}, S., \& {Feldman}, P.~D.
  2007, \apj, 668, 1174


\bibitem[Gagn{\'e} et al.(2014)]{gagne2014} Gagn{\'e}, J., 
Faherty, J.~K., Cruz, K., et al.\ 2014, \apjl, 785, L14 


\bibitem[{{Greaves} {et~al.}(2000){Greaves}, {Coulson}, \&
  {Holland}}]{greaves2000a}
{Greaves}, J.~S., {Coulson}, I.~M., \& {Holland}, W.~S. 2000, \mnras, 312, L1

\bibitem[{{Griffin} {et~al.}(2010){Griffin}, {Abergel}, {Abreu}, {Ade},
  {Andr{\'e}}, {Augueres}, {Babbedge}, {Bae}, {Baillie}, {Baluteau}, {Barlow},
  {Bendo}, {Benielli}, {Bock}, {Bonhomme}, {Brisbin}, {Brockley-Blatt},
  {Caldwell}, {Cara}, {Castro-Rodriguez}, {Cerulli}, {Chanial}, {Chen},
  {Clark}, {Clements}, {Clerc}, {Coker}, {Communal}, {Conversi}, {Cox},
  {Crumb}, {Cunningham}, {Daly}, {Davis}, {de Antoni}, {Delderfield}, {Devin},
  {di Giorgio}, {Didschuns}, {Dohlen}, {Donati}, {Dowell}, {Dowell}, {Duband},
  {Dumaye}, {Emery}, {Ferlet}, {Ferrand}, {Fontignie}, {Fox}, {Franceschini},
  {Frerking}, {Fulton}, {Garcia}, {Gastaud}, {Gear}, {Glenn}, {Goizel},
  {Griffin}, {Grundy}, {Guest}, {Guillemet}, {Hargrave}, {Harwit}, {Hastings},
  {Hatziminaoglou}, {Herman}, {Hinde}, {Hristov}, {Huang}, {Imhof}, {Isaak},
  {Israelsson}, {Ivison}, {Jennings}, {Kiernan}, {King}, {Lange}, {Latter},
  {Laurent}, {Laurent}, {Leeks}, {Lellouch}, {Levenson}, {Li}, {Li},
  {Lilienthal}, {Lim}, {Liu}, {Lu}, {Madden}, {Mainetti}, {Marliani}, {McKay},
  {Mercier}, {Molinari}, {Morris}, {Moseley}, {Mulder}, {Mur}, {Naylor},
  {Nguyen}, {O'Halloran}, {Oliver}, {Olofsson}, {Olofsson}, {Orfei}, {Page},
  {Pain}, {Panuzzo}, {Papageorgiou}, {Parks}, {Parr-Burman}, {Pearce},
  {Pearson}, {P{\'e}rez-Fournon}, {Pinsard}, {Pisano}, {Podosek}, {Pohlen},
  {Polehampton}, {Pouliquen}, {Rigopoulou}, {Rizzo}, {Roseboom}, {Roussel},
  {Rowan-Robinson}, {Rownd}, {Saraceno}, {Sauvage}, {Savage}, {Savini},
  {Sawyer}, {Scharmberg}, {Schmitt}, {Schneider}, {Schulz}, {Schwartz},
  {Shafer}, {Shupe}, {Sibthorpe}, {Sidher}, {Smith}, {Smith}, {Smith},
  {Spencer}, {Stobie}, {Sudiwala}, {Sukhatme}, {Surace}, {Stevens}, {Swinyard},
  {Trichas}, {Tourette}, {Triou}, {Tseng}, {Tucker}, {Turner}, {Vaccari},
  {Valtchanov}, {Vigroux}, {Virique}, {Voellmer}, {Walker}, {Ward}, {Waskett},
  {Weilert}, {Wesson}, {White}, {Whitehouse}, {Wilson}, {Winter}, {Woodcraft},
  {Wright}, {Xu}, {Zavagno}, {Zemcov}, {Zhang}, \& {Zonca}}]{griffin2010}
{Griffin}, M.~J., {Abergel}, A., {Abreu}, A., {et~al.} 2010, \aap, 518, L3

\bibitem[{{Grigorieva} {et~al.}(2007){Grigorieva}, {Th{\'e}bault},
  {Artymowicz}, \& {Brandeker}}]{grigorieva2007}
{Grigorieva}, A., {Th{\'e}bault}, P., {Artymowicz}, P., \& {Brandeker}, A.
  2007, \aap, 475, 755

\bibitem[{{G{\"u}sten} {et~al.}(2006){G{\"u}sten}, {Nyman}, {Schilke},
  {Menten}, {Cesarsky}, \& {Booth}}]{gusten2006}
{G{\"u}sten}, R., {Nyman}, L.~{\AA}., {Schilke}, P., {et~al.} 2006, \aap, 454,
  L13

\bibitem[{{Hales} {et~al.}(2014){Hales}, {De Gregorio-Monsalvo}, {Montesinos},
  {Casassus}, {Dent}, {Dougados}, {Eiroa}, {Hughes}, {Garay}, {Mardones},
  {M{\'e}nard}, {Palau}, {P{\'e}rez}, {Phillips}, {Torrelles}, \&
  {Wilner}}]{hales2014}
{Hales}, A.~S., {De Gregorio-Monsalvo}, I., {Montesinos}, B., {et~al.} 2014,
  \aj, 148, 47

\bibitem[{{Hauck} \& {Mermilliod}(1998)}]{hauck1998}
{Hauck}, B., \& {Mermilliod}, M. 1998, \aaps, 129, 431

\bibitem[{{Hobbs} {et~al.}(1985){Hobbs}, {Vidal-Madjar}, {Ferlet}, {Albert}, \&
  {Gry}}]{hobbs1985}
{Hobbs}, L.~M., {Vidal-Madjar}, A., {Ferlet}, R., {Albert}, C.~E., \& {Gry}, C.
  1985, \apjl, 293, L29

\bibitem[{{H{\o}g} {et~al.}(2000){H{\o}g}, {Fabricius}, {Makarov}, {Bastian},
  {Schwekendiek}, {Wicenec}, {Urban}, {Corbin}, \& {Wycoff}}]{hog2000}
{H{\o}g}, E., {Fabricius}, C., {Makarov}, V.~V., {et~al.} 2000, \aap, 357, 367

\bibitem[{{Houck} {et~al.}(2004){Houck}, {Roellig}, {van Cleve}, {Forrest},
  {Herter}, {Lawrence}, {Matthews}, {Reitsema}, {Soifer}, {Watson}, {Weedman},
  {Huisjen}, {Troeltzsch}, {Barry}, {Bernard-Salas}, {Blacken}, {Brandl},
  {Charmandaris}, {Devost}, {Gull}, {Hall}, {Henderson}, {Higdon}, {Pirger},
  {Schoenwald}, {Sloan}, {Uchida}, {Appleton}, {Armus}, {Burgdorf},
  {Fajardo-Acosta}, {Grillmair}, {Ingalls}, {Morris}, \& {Teplitz}}]{houck2004}
{Houck}, J.~R., {Roellig}, T.~L., {van Cleve}, J., {et~al.} 2004, \apjs, 154,
  18

\bibitem[{{Houk}(1982)}]{houk1982}
{Houk}, N. 1982, {Michigan Catalogue of Two-dimensional Spectral Types for the
  HD stars. Volume\_3. Declinations -40\fdg0 to -26\fdg0.}

\bibitem[{{Hung} {et~al.}(2015){Hung}, {Fitzgerald}, {Chen}, {Mittal}, {Kalas},
  \& {Graham}}]{hung2015}
{Hung}, L.-W., {Fitzgerald}, M.~P., {Chen}, C.~H., {et~al.} 2015, \apj, 802,
  138

\bibitem[{{Ishihara} {et~al.}(2010){Ishihara}, {Onaka}, {Kataza}, {Salama},
  {Alfageme}, {Cassatella}, {Cox}, {Garc{\'{\i}}a-Lario}, {Stephenson},
  {Cohen}, {Fujishiro}, {Fujiwara}, {Hasegawa}, {Ita}, {Kim}, {Matsuhara},
  {Murakami}, {M{\"u}ller}, {Nakagawa}, {Ohyama}, {Oyabu}, {Pyo}, {Sakon},
  {Shibai}, {Takita}, {Tanab{\'e}}, {Uemizu}, {Ueno}, {Usui}, {Wada},
  {Watarai}, {Yamamura}, \& {Yamauchi}}]{ishihara2010}
{Ishihara}, D., {Onaka}, T., {Kataza}, H., {et~al.} 2010, \aap, 514, A1

\bibitem[{{Johnson} {et~al.}(2012){Johnson}, {Lisse}, {Chen}, {Melosh},
  {Wyatt}, {Thebault}, {Henning}, {Gaidos}, {Elkins-Tanton}, {Bridges}, \&
  {Morlok}}]{johnson2012}
{Johnson}, B.~C., {Lisse}, C.~M., {Chen}, C.~H., {et~al.} 2012, \apj, 761, 45

\bibitem[{{Kamp} \& {van Zadelhoff}(2001)}]{kamp2001}
{Kamp}, I., \& {van Zadelhoff}, G.-J. 2001, \aap, 373, 641

\bibitem[{{Kastner} {et~al.}(2010){Kastner}, {Hily-Blant}, {Sacco},
  {Forveille}, \& {Zuckerman}}]{kastner2010}
{Kastner}, J.~H., {Hily-Blant}, P., {Sacco}, G.~G., {Forveille}, T., \&
  {Zuckerman}, B. 2010, \apjl, 723, L248

\bibitem[{{Kaufer} {et~al.}(1999){Kaufer}, {Stahl}, {Tubbesing},
  {N{\o}rregaard}, {Avila}, {Francois}, {Pasquini}, \& {Pizzella}}]{kaufer1999}
{Kaufer}, A., {Stahl}, O., {Tubbesing}, S., {et~al.} 1999, The Messenger, 95, 8

\bibitem[{{Kennedy} \& {Wyatt}(2014)}]{kennedy2014}
{Kennedy}, G.~M., \& {Wyatt}, M.~C. 2014, \mnras, 444, 3164

\bibitem[{{Kiefer} {et~al.}(2014){Kiefer}, {Lecavelier des Etangs}, {Augereau},
  {Vidal-Madjar}, {Lagrange}, \& {Beust}}]{kiefer2014}
{Kiefer}, F., {Lecavelier des Etangs}, A., {Augereau}, J.-C., {et~al.} 2014,
  \aap, 561, L10

\bibitem[{{Klein} {et~al.}(2012){Klein}, {Hochg{\"u}rtel}, {Kr{\"a}mer},
  {Bell}, {Meyer}, \& {G{\"u}sten}}]{klein2012}
{Klein}, B., {Hochg{\"u}rtel}, S., {Kr{\"a}mer}, I., {et~al.} 2012, \aap, 542,
  L3

\bibitem[{{Klein} {et~al.}(2014){Klein}, {Ciechanowicz}, {Leinz}, {Heyminck},
  {Gusten}, {Kasemann}, {Wunsch}, {Maier}, \& {Sekimoto}}]{klein2014}
{Klein}, T., {Ciechanowicz}, M., {Leinz}, C., {et~al.} 2014, IEEE Transactions
  on Terahertz Science and Technology, 4, 588

\bibitem[{{K{\'o}sp{\'a}l} {et~al.}(2013){K{\'o}sp{\'a}l}, {Mo{\'o}r},
  {Juh{\'a}sz}, {{\'A}brah{\'a}m}, {Apai}, {Csengeri}, {Grady}, {Henning},
  {Hughes}, {Kiss}, {Pascucci}, \& {Schmalzl}}]{kospal2013}
{K{\'o}sp{\'a}l}, {\'A}., {Mo{\'o}r}, A., {Juh{\'a}sz}, A., {et~al.} 2013,
  \apj, 776, 77

\bibitem[{{Kouwenhoven} {et~al.}(2005){Kouwenhoven}, {Brown}, {Zinnecker},
  {Kaper}, \& {Portegies Zwart}}]{kouwenhoven2005}
{Kouwenhoven}, M.~B.~N., {Brown}, A.~G.~A., {Zinnecker}, H., {Kaper}, L., \&
  {Portegies Zwart}, S.~F. 2005, \aap, 430, 137

\bibitem[{{Krivov}(2010)}]{krivov2010}
{Krivov}, A.~V. 2010, Research in Astronomy and Astrophysics, 10, 383


\bibitem[Lebouteiller et al.(2011)]{lebouteiller2011} Lebouteiller, V., 
Barry, D.~J., Spoon, H.~W.~W., et al.\ 2011, \apjs, 196, 8 

\bibitem[{{Lee}(1984)}]{lee1984}
{Lee}, L.~C. 1984, \apj, 282, 172

\bibitem[Lequeux(2005)]{lequeux2005} Lequeux, J.\ 2005, The 
interstellar medium, Translation from the French language edition of: Le 
Milieu Interstellaire by James Lequeux, EDP Sciences, 2003 Edited by 
J.~Lequeux.~ Astronomy and astrophysics library, Berlin: Springer, 2005


\bibitem[Mamajek \& Bell(2014)]{mamajek2014} Mamajek, E.~E., \& Bell, C.~P.~M.\ 2014, \mnras, 445, 2169 


\bibitem[{{Markwardt}(2009)}]{markwardt2009}
{Markwardt}, C.~B. 2009, in Astronomical Society of the Pacific Conference
  Series, Vol. 411, Astronomical Data Analysis Software and Systems XVIII, ed.
  D.~A. {Bohlender}, D.~{Durand}, \& P.~{Dowler}, 251

\bibitem[{{Matr{\`a}} {et~al.}(2015){Matr{\`a}}, {Pani{\'c}}, {Wyatt}, \&
  {Dent}}]{matra2015}
{Matr{\`a}}, L., {Pani{\'c}}, O., {Wyatt}, M.~C., \& {Dent}, W.~R.~F. 2015,
  \mnras, 447, 3936

\bibitem[{{Matthews} {et~al.}(2014){Matthews}, {Krivov}, {Wyatt}, {Bryden}, \&
  {Eiroa}}]{matthews2014b}
{Matthews}, B.~C., {Krivov}, A.~V., {Wyatt}, M.~C., {Bryden}, G., \& {Eiroa},
  C. 2014, Protostars and Planets VI, 521

\bibitem[{{Meeus} {et~al.}(2012){Meeus}, {Montesinos}, {Mendigut{\'{\i}}a},
  {Kamp}, {Thi}, {Eiroa}, {Grady}, {Mathews}, {Sandell}, {Martin-Za{\"i}di},
  {Brittain}, {Dent}, {Howard}, {M{\'e}nard}, {Pinte}, {Roberge},
  {Vandenbussche}, \& {Williams}}]{meeus2012}
{Meeus}, G., {Montesinos}, B., {Mendigut{\'{\i}}a}, I., {et~al.} 2012, \aap,
  544, A78

\bibitem[{{Melis} {et~al.}(2013){Melis}, {Zuckerman}, {Rhee}, {Song}, {Murphy},
  \& {Bessell}}]{melis2013}
{Melis}, C., {Zuckerman}, B., {Rhee}, J.~H., {et~al.} 2013, \apj, 778, 12

\bibitem[{{Mendigut{\'{\i}}a} {et~al.}(2011){Mendigut{\'{\i}}a}, {Calvet},
  {Montesinos}, {Mora}, {Muzerolle}, {Eiroa}, {Oudmaijer}, \&
  {Mer{\'{\i}}n}}]{mendigutia2011}
{Mendigut{\'{\i}}a}, I., {Calvet}, N., {Montesinos}, B., {et~al.} 2011, \aap,
  535, A99

\bibitem[{{Montgomery} \& {Welsh}(2012)}]{montgomery2012}
{Montgomery}, S.~L., \& {Welsh}, B.~Y. 2012, \pasp, 124, 1042

\bibitem[{{Mo{\'o}r} {et~al.}(2006){Mo{\'o}r}, {{\'A}brah{\'a}m}, {Derekas},
  {Kiss}, {Kiss}, {Apai}, {Grady}, \& {Henning}}]{moor2006}
{Mo{\'o}r}, A., {{\'A}brah{\'a}m}, P., {Derekas}, A., {et~al.} 2006, \apj, 644,
  525

\bibitem[{{Mo{\'o}r} {et~al.}(2011{\natexlab{a}}){Mo{\'o}r}, {{\'A}brah{\'a}m},
  {Juh{\'a}sz}, {Kiss}, {Pascucci}, {K{\'o}sp{\'a}l}, {Apai}, {Henning},
  {Csengeri}, \& {Grady}}]{moor2011b}
{Mo{\'o}r}, A., {{\'A}brah{\'a}m}, P., {Juh{\'a}sz}, A., {et~al.}
  2011{\natexlab{a}}, \apjl, 740, L7

\bibitem[{{Mo{\'o}r} {et~al.}(2011{\natexlab{b}}){Mo{\'o}r}, {Pascucci},
  {K{\'o}sp{\'a}l}, {{\'A}brah{\'a}m}, {Csengeri}, {Kiss}, {Apai}, {Grady},
  {Henning}, {Kiss}, {Bayliss}, {Juh{\'a}sz}, {Kov{\'a}cs}, \&
  {Szalai}}]{moor2011a}
{Mo{\'o}r}, A., {Pascucci}, I., {K{\'o}sp{\'a}l}, {\'A}., {et~al.}
  2011{\natexlab{b}}, \apjs, 193, 4

\bibitem[{{Mo{\'o}r} {et~al.}(2013){Mo{\'o}r}, {{\'A}brah{\'a}m},
  {K{\'o}sp{\'a}l}, {Szab{\'o}}, {Apai}, {Balog}, {Csengeri}, {Grady},
  {Henning}, {Juh{\'a}sz}, {Kiss}, {Pascucci}, {Szul{\'a}gyi}, \&
  {Vavrek}}]{moor2013a}
{Mo{\'o}r}, A., {{\'A}brah{\'a}m}, P., {K{\'o}sp{\'a}l}, {\'A}., {et~al.} 2013,
  \apjl, 775, L51

\bibitem[{{Mo{\'o}r} {et~al.}(2015){Mo{\'o}r}, {K{\'o}sp{\'a}l},
  {{\'A}brah{\'a}m}, {Apai}, {Balog}, {Grady}, {Henning}, {Juh{\'a}sz}, {Kiss},
  {Krivov}, {Pawellek}, \& {Szab{\'o}}}]{moor2015}
{Mo{\'o}r}, A., {K{\'o}sp{\'a}l}, {\'A}., {{\'A}brah{\'a}m}, P., {et~al.} 2015,
  \mnras, 447, 577

\bibitem[{{Morales} {et~al.}(2013){Morales}, {Bryden}, {Werner}, \&
  {Stapelfeldt}}]{morales2013}
{Morales}, F.~Y., {Bryden}, G., {Werner}, M.~W., \& {Stapelfeldt}, K.~R. 2013,
  \apj, 776, 111

\bibitem[{{Moshir}(1990)}]{moshir}
{Moshir}, M.~et al. 1990, in IRAS Faint Source Catalogue, version 2.0 (1990), 0

\bibitem[{{Mumma} \& {Charnley}(2011)}]{mumma2011}
{Mumma}, M.~J., \& {Charnley}, S.~B. 2011, \araa, 49, 471

\bibitem[{{Munari} {et~al.}(2005){Munari}, {Sordo}, {Castelli}, \&
  {Zwitter}}]{munari2005}
{Munari}, U., {Sordo}, R., {Castelli}, F., \& {Zwitter}, T. 2005, \aap, 442,
  1127

\bibitem[{{Muzerolle} {et~al.}(2004){Muzerolle}, {D'Alessio}, {Calvet}, \&
  {Hartmann}}]{muzerolle2004}
{Muzerolle}, J., {D'Alessio}, P., {Calvet}, N., \& {Hartmann}, L. 2004, \apj,
  617, 406

\bibitem[{{Najita} \& {Williams}(2005)}]{najita2005}
{Najita}, J., \& {Williams}, J.~P. 2005, \apj, 635, 625

\bibitem[{{Nielsen} {et~al.}(2013){Nielsen}, {Liu}, {Wahhaj}, {Biller},
  {Hayward}, {Close}, {Males}, {Skemer}, {Chun}, {Ftaclas}, {Alencar},
  {Artymowicz}, {Boss}, {Clarke}, {de Gouveia Dal Pino}, {Gregorio-Hetem},
  {Hartung}, {Ida}, {Kuchner}, {Lin}, {Reid}, {Shkolnik}, {Tecza}, {Thatte}, \&
  {Toomey}}]{nielsen2013}
{Nielsen}, E.~L., {Liu}, M.~C., {Wahhaj}, Z., {et~al.} 2013, \apj, 776, 4

\bibitem[{{Nilsson} {et~al.}(2010){Nilsson}, {Liseau}, {Brandeker}, {Olofsson},
  {Pilbratt}, {Risacher}, {Rodmann}, {Augereau}, {Bergman}, {Eiroa},
  {Fridlund}, {Th{\'e}bault}, \& {White}}]{nilsson2010}
{Nilsson}, R., {Liseau}, R., {Brandeker}, A., {et~al.} 2010, \aap, 518, A40

\bibitem[{{Olsen}(1984)}]{olsen1984}
{Olsen}, E.~H. 1984, \aaps, 57, 443

\bibitem[{{Ott}(2010)}]{ott2010}
{Ott}, S. 2010, in Astronomical Society of the Pacific Conference Series, Vol.
  434, Astronomical Data Analysis Software and Systems XIX, ed. Y.~{Mizumoto},
  K.-I. {Morita}, \& M.~{Ohishi}, 139

\bibitem[{{Pascucci} {et~al.}(2006){Pascucci}, {Gorti}, {Hollenbach}, {Najita},
  {Meyer}, {Carpenter}, {Hillenbrand}, {Herczeg}, {Padgett}, {Mamajek},
  {Silverstone}, {Schlingman}, {Kim}, {Stobie}, {Bouwman}, {Wolf}, {Rodmann},
  {Hines}, {Lunine}, \& {Malhotra}}]{pascucci2006}
{Pascucci}, I., {Gorti}, U., {Hollenbach}, D., {et~al.} 2006, \apj, 651, 1177

\bibitem[{{Patel} {et~al.}(2014){Patel}, {Metchev}, \& {Heinze}}]{patel2014}
{Patel}, R.~I., {Metchev}, S.~A., \& {Heinze}, A. 2014, \apjs, 212, 10

\bibitem[{{Pawellek} {et~al.}(2014){Pawellek}, {Krivov}, {Marshall},
  {Montesinos}, {{\'A}brah{\'a}m}, {Mo{\'o}r}, {Bryden}, \&
  {Eiroa}}]{pawellek2014}
{Pawellek}, N., {Krivov}, A.~V., {Marshall}, J.~P., {et~al.} 2014, \apj, 792,
  65

\bibitem[Pecaut et al.(2012)]{pecaut2011} Pecaut, M.~J., Mamajek, 
E.~E., \& Bubar, E.~J.\ 2012, \apj, 746, 154 

\bibitem[{{Pecaut} \& {Mamajek}(2013)}]{pecaut2013}
{Pecaut}, M.~J., \& {Mamajek}, E.~E. 2013, \apjs, 208, 9

\bibitem[{{Perryman}(1997)}]{perryman1997}
{Perryman}, M.~A.~C. 1997, in ESA Special Publication, Vol. 402, Hipparcos -
  Venice '97, ed. R.~M. {Bonnet}, E.~{H{\o}g}, P.~L. {Bernacca}, L.~{Emiliani},
  A.~{Blaauw}, C.~{Turon}, J.~{Kovalevsky}, L.~{Lindegren}, H.~{Hassan},
  M.~{Bouffard}, B.~{Strim}, D.~{Heger}, M.~A.~C. {Perryman}, \& L.~{Woltjer},
  1--4

\bibitem[{{Pilbratt} {et~al.}(2010){Pilbratt}, {Riedinger}, {Passvogel},
  {Crone}, {Doyle}, {Gageur}, {Heras}, {Jewell}, {Metcalfe}, {Ott}, \&
  {Schmidt}}]{pilbratt2010}
{Pilbratt}, G.~L., {Riedinger}, J.~R., {Passvogel}, T., {et~al.} 2010, \aap,
  518, L1

\bibitem[{{Poglitsch} {et~al.}(2010){Poglitsch}, {Waelkens}, {Geis},
  {Feuchtgruber}, {Vandenbussche}, {Rodriguez}, {Krause}, {Renotte}, {van
  Hoof}, {Saraceno}, {Cepa}, {Kerschbaum}, {Agn{\`e}se}, {Ali}, {Altieri},
  {Andreani}, {Augueres}, {Balog}, {Barl}, {Bauer}, {Belbachir}, {Benedettini},
  {Billot}, {Boulade}, {Bischof}, {Blommaert}, {Callut}, {Cara}, {Cerulli},
  {Cesarsky}, {Contursi}, {Creten}, {De Meester}, {Doublier}, {Doumayrou},
  {Duband}, {Exter}, {Genzel}, {Gillis}, {Gr{\"o}zinger}, {Henning},
  {Herreros}, {Huygen}, {Inguscio}, {Jakob}, {Jamar}, {Jean}, {de Jong},
  {Katterloher}, {Kiss}, {Klaas}, {Lemke}, {Lutz}, {Madden}, {Marquet},
  {Martignac}, {Mazy}, {Merken}, {Montfort}, {Morbidelli}, {M{\"u}ller},
  {Nielbock}, {Okumura}, {Orfei}, {Ottensamer}, {Pezzuto}, {Popesso},
  {Putzeys}, {Regibo}, {Reveret}, {Royer}, {Sauvage}, {Schreiber}, {Stegmaier},
  {Schmitt}, {Schubert}, {Sturm}, {Thiel}, {Tofani}, {Vavrek}, {Wetzstein},
  {Wieprecht}, \& {Wiezorrek}}]{poglitsch2010}
{Poglitsch}, A., {Waelkens}, C., {Geis}, N., {et~al.} 2010, \aap, 518, L2

\bibitem[{{Redfield}(2007)}]{redfield2007}
{Redfield}, S. 2007, \apjl, 656, L97

\bibitem[{{Redfield} {et~al.}(2007){Redfield}, {Kessler-Silacci}, \&
  {Cieza}}]{redfield2007b}
{Redfield}, S., {Kessler-Silacci}, J.~E., \& {Cieza}, L.~A. 2007, \apj, 661,
  944


\bibitem[{Regibo} (2012)]{regibo2012} Regibo, S. 2012, PhD Thesis, Data Reduction and Analysis Algorithms for the
Herschel Space Observatory, KUL Leuven


\bibitem[{{Rhee} {et~al.}(2007){Rhee}, {Song}, {Zuckerman}, \&
  {McElwain}}]{rhee2007}
{Rhee}, J.~H., {Song}, I., {Zuckerman}, B., \& {McElwain}, M. 2007, \apj, 660,
  1556

\bibitem[{{Ricci} {et~al.}(2015){Ricci}, {Carpenter}, {Fu}, {Hughes}, {Corder},
  \& {Isella}}]{ricci2015}
{Ricci}, L., {Carpenter}, J.~M., {Fu}, B., {et~al.} 2015, \apj, 798, 124

\bibitem[{{Rigliaco} {et~al.}(2015){Rigliaco}, {Pascucci}, {Duchene},
  {Edwards}, {Ardila}, {Grady}, {Mendigut{\'{\i}}a}, {Montesinos}, {Mulders},
  {Najita}, {Carpenter}, {Furlan}, {Gorti}, {Meijerink}, \&
  {Meyer}}]{rigliaco2015}
{Rigliaco}, E., {Pascucci}, I., {Duchene}, G., {et~al.} 2015, \apj, 801, 31

\bibitem[{{Riviere-Marichalar} {et~al.}(2012){Riviere-Marichalar}, {Barrado},
  {Augereau}, {Thi}, {Roberge}, {Eiroa}, {Montesinos}, {Meeus}, {Howard},
  {Sandell}, {Duch{\^e}ne}, {Dent}, {Lebreton}, {Mendigut{\'{\i}}a},
  {Hu{\'e}lamo}, {M{\'e}nard}, \& {Pinte}}]{rm2012}
{Riviere-Marichalar}, P., {Barrado}, D., {Augereau}, J.-C., {et~al.} 2012,
  \aap, 546, L8

\bibitem[{{Riviere-Marichalar} {et~al.}(2014){Riviere-Marichalar}, {Barrado},
  {Montesinos}, {Duch{\^e}ne}, {Bouy}, {Pinte}, {Menard}, {Donaldson}, {Eiroa},
  {Krivov}, {Kamp}, {Mendigut{\'{\i}}a}, {Dent}, \& {Lillo-Box}}]{rm2014}
{Riviere-Marichalar}, P., {Barrado}, D., {Montesinos}, B., {et~al.} 2014, \aap,
  565, A68

\bibitem[{{Roberge} {et~al.}(2006){Roberge}, {Feldman}, {Weinberger},
  {Deleuil}, \& {Bouret}}]{roberge2006}
{Roberge}, A., {Feldman}, P.~D., {Weinberger}, A.~J., {Deleuil}, M., \&
  {Bouret}, J.-C. 2006, \nat, 441, 724

\bibitem[{{Roberge} {et~al.}(2014){Roberge}, {Welsh}, {Kamp}, {Weinberger}, \&
  {Grady}}]{roberge2014}
{Roberge}, A., {Welsh}, B.~Y., {Kamp}, I., {Weinberger}, A.~J., \& {Grady},
  C.~A. 2014, \apjl, 796, L11

\bibitem[{{Roberge} {et~al.}(2013){Roberge}, {Kamp}, {Montesinos}, {Dent},
  {Meeus}, {Donaldson}, {Olofsson}, {Mo{\'o}r}, {Augereau}, {Howard}, {Eiroa},
  {Thi}, {Ardila}, {Sandell}, \& {Woitke}}]{roberge2013}
{Roberge}, A., {Kamp}, I., {Montesinos}, B., {et~al.} 2013, \apj, 771, 69

\bibitem[{{Rodigas} {et~al.}(2014){Rodigas}, {Debes}, {Hinz}, {Mamajek},
  {Pecaut}, {Currie}, {Bailey}, {Defrere}, {De Rosa}, {Hill}, {Leisenring},
  {Schneider}, {Skemer}, {Skrutskie}, {Vaitheeswaran}, \&
  {Ward-Duong}}]{rodigas2014}
{Rodigas}, T.~J., {Debes}, J.~H., {Hinz}, P.~M., {et~al.} 2014, \apj, 783, 21

\bibitem[{{Schlafly} \& {Finkbeiner}(2011)}]{schlafly2011}
{Schlafly}, E.~F., \& {Finkbeiner}, D.~P. 2011, \apj, 737, 103

\bibitem[{{Sheret} {et~al.}(2004){Sheret}, {Dent}, \& {Wyatt}}]{sheret2004}
{Sheret}, I., {Dent}, W.~R.~F., \& {Wyatt}, M.~C. 2004, \mnras, 348, 1282

\bibitem[{{Siess} {et~al.}(2000){Siess}, {Dufour}, \& {Forestini}}]{siess2000}
{Siess}, L., {Dufour}, E., \& {Forestini}, M. 2000, \aap, 358, 593

\bibitem[{{Slawson} {et~al.}(1992){Slawson}, {Hill}, \&
  {Landstreet}}]{slawson1992}
{Slawson}, R.~W., {Hill}, R.~J., \& {Landstreet}, J.~D. 1992, \apjs, 82, 117

\bibitem[{{Slettebak}(1975)}]{slettebak1975}
{Slettebak}, A. 1975, \apj, 197, 137


\bibitem[Torres et al.(2008)]{torres2008} Torres, C.~A.~O., Quast, 
G.~R., Melo, C.~H.~F., \& Sterzik, M.~F.\ 2008, Handbook of Star Forming Regions, Volume II: The Southern Sky ASP
Monograph Publications, Vol. 5. Edited by Bo Reipurth, p.757 


\bibitem[{{van Leeuwen}(2007)}]{vanleeuwen2007}
{van Leeuwen}, F., ed. 2007, Astrophysics and Space Science Library, Vol. 350,
  {Hipparcos, the New Reduction of the Raw Data}

\bibitem[{{Vassilev} {et~al.}(2008){Vassilev}, {Meledin}, {Lapkin}, {Belitsky},
  {Nystr{\"o}m}, {Henke}, {Pavolotsky}, {Monje}, {Risacher}, {Olberg},
  {Strandberg}, {Sundin}, {Fredrixon}, {Ferm}, {Desmaris}, {Dochev},
  {Pantaleev}, {Bergman}, \& {Olofsson}}]{vassilev2008}
{Vassilev}, V., {Meledin}, D., {Lapkin}, I., {et~al.} 2008, \aap, 490, 1157

\bibitem[{{Vican} \& {Schneider}(2014)}]{vican2014}
{Vican}, L., \& {Schneider}, A. 2014, \apj, 780, 154

\bibitem[{{Visser} {et~al.}(2009){Visser}, {van Dishoeck}, \&
  {Black}}]{visser2009}
{Visser}, R., {van Dishoeck}, E.~F., \& {Black}, J.~H. 2009, \aap, 503, 323

\bibitem[{{Wahhaj} {et~al.}(2013){Wahhaj}, {Liu}, {Nielsen}, {Biller},
  {Hayward}, {Close}, {Males}, {Skemer}, {Ftaclas}, {Chun}, {Thatte}, {Tecza},
  {Shkolnik}, {Kuchner}, {Reid}, {de Gouveia Dal Pino}, {Alencar},
  {Gregorio-Hetem}, {Boss}, {Lin}, \& {Toomey}}]{wahhaj2013}
{Wahhaj}, Z., {Liu}, M.~C., {Nielsen}, E.~L., {et~al.} 2013, \apj, 773, 179

\bibitem[Weinberger et al.(2000)]{weinberger2000} Weinberger, A.~J., 
Rich, R.~M., Becklin, E.~E., Zuckerman, B., 
\& Matthews, K.\ 2000, \apj, 544, 937 



\bibitem[{{Welsh} \& {Montgomery}(2013)}]{welsh2013}
{Welsh}, B.~Y., \& {Montgomery}, S. 2013, \pasp, 125, 759

\bibitem[{{Werner} {et~al.}(2004){Werner}, {Roellig}, {Low}, {Rieke}, {Rieke},
  {Hoffmann}, {Young}, {Houck}, {Brandl}, {Fazio}, {Hora}, {Gehrz}, {Helou},
  {Soifer}, {Stauffer}, {Keene}, {Eisenhardt}, {Gallagher}, {Gautier}, {Irace},
  {Lawrence}, {Simmons}, {Van Cleve}, {Jura}, {Wright}, \&
  {Cruikshank}}]{werner2004}
{Werner}, M.~W., {Roellig}, T.~L., {Low}, F.~J., {et~al.} 2004, \apjs, 154, 1

\bibitem[{{Williams} \& {Cieza}(2011)}]{williams2011}
{Williams}, J.~P., \& {Cieza}, L.~A. 2011, \araa, 49, 67

\bibitem[{{Wilson} \& {Bell}(2002)}]{wilson2002}
{Wilson}, N.~J., \& {Bell}, K.~L. 2002, \mnras, 337, 1027

\bibitem[{{Wright} {et~al.}(2010){Wright}, {Eisenhardt}, {Mainzer}, {Ressler},
  {Cutri}, {Jarrett}, {Kirkpatrick}, {Padgett}, {McMillan}, {Skrutskie},
  {Stanford}, {Cohen}, {Walker}, {Mather}, {Leisawitz}, {Gautier}, {McLean},
  {Benford}, {Lonsdale}, {Blain}, {Mendez}, {Irace}, {Duval}, {Liu}, {Royer},
  {Heinrichsen}, {Howard}, {Shannon}, {Kendall}, {Walsh}, {Larsen}, {Cardon},
  {Schick}, {Schwalm}, {Abid}, {Fabinsky}, {Naes}, \& {Tsai}}]{wright2010}
{Wright}, E.~L., {Eisenhardt}, P.~R.~M., {Mainzer}, A.~K., {et~al.} 2010, \aj,
  140, 1868

\bibitem[{{Wyatt}(2008)}]{wyatt2008}
{Wyatt}, M.~C. 2008, \araa, 46, 339

\bibitem[{{Wyatt} {et~al.}(2015){Wyatt}, {Pani{\'c}}, {Kennedy}, \&
  {Matr{\`a}}}]{wyatt2015}
{Wyatt}, M.~C., {Pani{\'c}}, O., {Kennedy}, G.~M., \& {Matr{\`a}}, L. 2015,
  \apss, 357, 103

\bibitem[{{Yang} {et~al.}(2013){Yang}, {Stancil}, {Balakrishnan}, {Forrey}, \&
  {Bowman}}]{yang2013}
{Yang}, B., {Stancil}, P.~C., {Balakrishnan}, N., {Forrey}, R.~C., \& {Bowman},
  J.~M. 2013, \apj, 771, 49

\bibitem[{{Zuckerman} {et~al.}(1995){Zuckerman}, {Forveille}, \&
  {Kastner}}]{zuckerman1995}
{Zuckerman}, B., {Forveille}, T., \& {Kastner}, J.~H. 1995, \nat, 373, 494

\bibitem[{{Zuckerman} \& {Song}(2012)}]{zuckerman2012}
{Zuckerman}, B., \& {Song}, I. 2012, \apj, 758, 77

\end{thebibliography}

\clearpage

\setlength{\tabcolsep}{4pt}
\begin{deluxetable}{lcccccccccc}                                                                                                                                                                            
\tabletypesize{\scriptsize}                                                                                                                                                                                  
\tablecaption{Stellar properties and CO observations for our targets. \label{basictable}}                                                                                                                    
\tablewidth{0pt}                                                                                                                                                                                             
\tablecolumns{11}                                                                                                                                                                                            
\tablehead{ \colhead{ID} & \colhead{SpT} &                                                                                                                                                                   
\colhead{Dist.} &  \colhead{Age} &  \colhead{Memb.} &                                                                                                                                                        
\colhead{$L_{\rm IR} / L_{\rm bol}$} &  \colhead{Prog.} &                                                                                                                                                    
\colhead{$S_{\rm CO(2-1)}$} & \colhead{$S_{\rm CO(3-2)}$} &                                                                                                                                                  
\colhead{$S_{\rm CO(4-3)}$} & \colhead{$M_{\rm CO}$} \\                                                                                                                                                      
\colhead{} & \colhead{} & \colhead{(pc)} &  \colhead{(Myr)} &                                                                                                                                                
\colhead{} & \colhead{} &  \colhead{} &                                                                                                                                                                      
\colhead{(Jy\,km\,s$^{-1}$)} & \colhead{(Jy\,km\,s$^{-1}$)} &                                                                                                                                                
\colhead{(Jy\,km\,s$^{-1}$)} & \colhead{($M_\oplus$)} \\                                                                                                                                                     
\colhead{(1)} & \colhead{(2)} & \colhead{(3)} &  \colhead{(4)} &                                                                                                                                             
\colhead{(5)} & \colhead{(6)} &  \colhead{(7)} &                                                                                                                                                             
\colhead{(8)} & \colhead{(9)} &                                                                                                                                                                              
\colhead{(10)} & \colhead{(11)} \\                                                                                                                                                                           
 }                                                                                                                                                                                                           
\startdata                                                                                                                                                                                                   
\multicolumn{11}{c}{Young debris disks with $f_d > 10^{-4}$} \\
                      HD 16743 & F0/F2III &   58.9 &      30 (2) &        - &   5.1$\times$10$^{-4}$ (3) &     2 &        \ldots &       $<$1.24 &        \ldots &          $<$4.6$\times10^{-5}$ \\
     HD 38206\tablenotemark{b} &      A0V &   75.1 &         30 &      COL &   1.4$\times$10$^{-4}$ (1) &     2 &        \ldots &       $<$2.40 &        \ldots &          $<$1.4$\times10^{-4}$ \\
     HD 95086\tablenotemark{a} &    A8III &   90.4 &         17 &      LCC &   1.7$\times$10$^{-3}$ (3) &     2 &        \ldots &       $<$1.95 &        \ldots &          $<$1.7$\times10^{-4}$ \\
                     HD 121191 &   A5IV/V &  130.0 &         17 &      LCC &   4.9$\times$10$^{-3}$ (2) &     4 &        \ldots &       $<$1.47 &        \ldots &          $<$2.6$\times10^{-4}$ \\
                     HD 131488 &      A1V &  150.0 &         16 &      UCL &   6.0$\times$10$^{-3}$ (2) &     4 &        \ldots &       $<$1.39 &        \ldots &          $<$3.3$\times10^{-4}$ \\
    HD 131835\tablenotemark{b} &     A2IV\tablenotemark{c} &  122.7 &         16 &      UCL &   3.0$\times$10$^{-3}$ (5) & 1,2,3 &   1.60$\pm$0.78   & 2.74$\pm$0.55 &    4.46$\pm$2.99 &
    $>$5.2$\times10^{-4}$\tablenotemark{e} \\
    HD 181296\tablenotemark{b} &      A0V &   48.2 &         23 &     BPMG &   2.4$\times$10$^{-4}$ (1) &     4 &        \ldots &       $<$1.38 &       $<$5.36 &          $<$3.4$\times10^{-5}$ \\
\multicolumn{11}{c}{Older A-type stars with extended, cold debris disks} \\
                      HD 10939 &      A1V &   62.0 &     346 (4) &        - &   1.2$\times$10$^{-4}$ (3) &     2 &        \ldots &       $<$1.37 &        \ldots &          $<$5.6$\times10^{-5}$ \\
     HD 17848\tablenotemark{b} &      A2V &   50.5 &     372 (4) &        - &   7.6$\times$10$^{-5}$ (3) &     2 &        \ldots &       $<$1.42 &        \ldots &          $<$3.8$\times10^{-5}$ \\
                     HD 161868 &      A0V &   31.5 &     450 (3) &        - &   1.3$\times$10$^{-4}$ (3) &     2 &        \ldots &       $<$1.82 &        \ldots &          $<$1.9$\times10^{-5}$ \\
                     HD 182681 &   B8/B9V &   69.9 &     144 (4) &        - &   2.8$\times$10$^{-4}$ (3) &     2 &        \ldots &       $<$2.15 &        \ldots &          $<$1.1$\times10^{-4}$ \\
\multicolumn{11}{c}{Debris disks around G--K-type Sun-like stars} \\
       HD 105\tablenotemark{b} &      G0V &   39.4 &         30 &      THA &   1.9$\times$10$^{-4}$ (1) &     3 &        \ldots &       $<$1.42 &       $<$6.08 &          $<$2.4$\times10^{-5}$ \\
       HD 377\tablenotemark{b} &      G2V &   39.1 &     183 (5)\tablenotemark{d} &        - &   2.6$\times$10$^{-4}$(1) &     3 &        \ldots &       $<$1.03 &       $<$2.41 &          $<$1.7$\times10^{-5}$ \\
     HD 61005\tablenotemark{b} &   G3/G5V &   35.3 &         40 &      ARG &   1.6$\times$10$^{-3}$ (1) &     3 &        \ldots &       $<$1.27 &       $<$4.37 &          $<$1.7$\times10^{-5}$ \\
                      HD 92945 &      K1V &   21.4 &     211 (5)\tablenotemark{d} &        - &   4.9$\times$10$^{-4}$(1) &     3 &        \ldots &       $<$1.33 &       $<$3.86 &          $<$0.7$\times10^{-5}$ \\
                     HD 160305 &   F8/G0V &   72.5 &         23 &     BPMG &   8.6$\times$10$^{-5}$ (4) &     3 &        \ldots &       $<$1.61 &       $<$5.54 &          $<$9.0$\times10^{-5}$ \\
                     HD 202917 &      G5V &   43.0 &         30 &      THA &   2.8$\times$10$^{-4}$ (1) &     3 &        \ldots &       $<$1.46 &       $<$7.22 &          $<$2.9$\times10^{-5}$ \\
\multicolumn{11}{c}{A-type stars with debris disks and variable Ca absorption line} \\
                     HD 110411 &      A0V &   36.3 &      90 (4) &        - &   5.9$\times$10$^{-5}$ (1) &     5 &       $<$0.20 &        \ldots &        \ldots &          $<$0.6$\times10^{-5}$ \\
                     HD 182919 &      A0V &   72.9 &     198 (1) &        - &   3.4$\times$10$^{-5}$ (1) &     5 &       $<$0.16 &        \ldots &        \ldots &          $<$2.0$\times10^{-5}$ \\
                     HD 183324 &      A0V &   61.2 &     140 (1) &        - &   2.2$\times$10$^{-5}$ (1) &     5 &       $<$0.16 &        \ldots &        \ldots &          $<$1.4$\times10^{-5}$ \\
\enddata                                                                                                                                                                                                
\tablenotetext{a}{CO observational data regarding this object are already                                                                                                                               
published in \citet{moor2013a}.}                                                                                                                                                                        
\tablenotetext{b}{HD\,17848, HD\,105, HD\,377, HD\,61005 and HD\,181296                                                                                                                                 
were also observed at the CO (3--2) transition                                                                                                                                                          
by \citet{hales2014} using the APEX and ASTE radio telescopes, for HD\,377                                                                                                                              
observation at 2--1 transition is also available                                                                                                                                                        
in \citet{pascucci2006}.                                                                                                                                                                                
HD\,38206 and HD\,131835 were searched for gas at 2--1 transition of CO                                                                                                                                 
\citep{kastner2010,zuckerman2012}. All of these previous                                                                                                                                                
measurements produced non-detections for these targets.
}
\tablenotetext{c}{The effective temperature of this star estimated by us (see Sect.~\ref{stellarprops}) 
is rather more consistent with a spectral type of A4. }
\tablenotetext{d}{\citet{vican2014} provided two different age estimates for these 
targets. Here we quoted the average of the two values.} 
\tablenotetext{e}{Derived from an LTE model (see Sect.~\ref{gasparams}). }                                                                                                                                      
\tablecomments{                                                                                                                                                                                         
 Col.(1): Identification.                                                                                                                                                                               
 Col.(2): Spectral type.                                                                                                                                                                                
 Col.(3): Distance. Based on {\sl Hipparcos} trigonometric parallax when it is available.                                                                                                               
 For HD\,121191 and HD\,131488 distances were taken from \citet{melis2013}.                                                                                                                             
 Col.(4): Stellar age. For members of young moving groups and associations                                                                                                                              
 we adopted the age of the corresponding group. For the other objects                                                                                                                                   
 the age information were taken from the literature (references are in brackets): 1) \citet{chen2014};                                                                                                  
 2) \citet{moor2011a}; 3) \citet{moor2015}; 4) \citet{nielsen2013}; 5) \citet{vican2014}.                                                                                                               
 Col.(5): Membership status of the star. ARG: the Argus moving group; BPMG: the $\beta$\,Pic moving group;                                                                                              
 COL: the Columba moving group; LCC: the Lower Centaurus Crux association;                                                                                                                              
 THA: the Tucana-Horologium association; UCL: the Upper Centaurus Lupus association.                                                                                                                    
 Col.(6): Fractional dust luminosity. Apart from HD\,131835 the data were taken from the literature:                                                                                                    
 1) \citet{chen2014};                                                                                                                                                                                   
 2) \citet{melis2013}; 3) \citet{moor2015}; 4) \citet{patel2014}; 5) this work.                                                                                                                         
 Col.(7): Observation projects. 1) - E-083.C-0303; 2) - M-087.F-0001; 3) - M-092.F-0012;                                                                                                                
 4) - M-093.F-0010; 5) - IRAM-172-13.                                                                                                                                                                   
 Col.(8--10): integrated line fluxes for CO $J$=2$-$1, $J$=3$-$2, $J$=4$-$3 (see Sect.~\ref{outcome}).                                                                                                  
 Col.(11): estimated mass of the CO gas.                                                                                                                                                                
}                                                                                                                                                                                                       
\end{deluxetable}                        
                                                                                                                                                                 
\begin{deluxetable}{lcccc}
\tabletypesize{\scriptsize}
\tablecaption{Observation  log. \label{obslog}}
\tablewidth{0pt}                                                                                                                                                                    
\tablecolumns{6}
\tablehead{\colhead{project ID} & \colhead{telescope} & \colhead{receiver} & \colhead{backend} & \colhead{obs.  period}  \\
}
\startdata
E-083.C-0303 & APEX & SHeFI/APEX2 & FFTS        & 2009-05-18              \\ 
M-087.F-0001-2011 & APEX & SHeFI/APEX2 & FFTS,XFFTS  & 2011-04-14--2011-12-16   \\ 
             & APEX & SHeFI/APEX1 & XFFTS       &                         \\ 
M-092.F-0012-2013 & APEX & FLASH+       & XFFTS       & 2013-09-29--2013-10-17  \\
M-093.F-0010-2014 & APEX & FLASH+       & XFFTS       & 2014-04-23--2014-04-27   \\
IRAM-172-13  & IRAM & EMIR        & FTS         & 2014-01-08              \\
\enddata 
\end{deluxetable} 


\begin{deluxetable}{ccccc}                                                      
\tabletypesize{\scriptsize}                                                     
\tablecaption{Measured and predicted photospheric IR flux densities for HD\,131835 \label{phottable}}
\tablewidth{0pt}                                                                
\tablecolumns{5}                                                                
\tablehead{ \colhead{$\lambda$} & \colhead{Meas. flux$^{a}$} &                  
\colhead{Instr.} &  \colhead{Pred. flux} &  \colhead{Reference} \\              
\colhead{[{\micron}]} & \colhead{[mJy]} &                                       
\colhead{} &  \colhead{[mJy]} &  \colhead{}                                     
}                                                                               
\startdata                                                                      
        3.35 &         297.0$\pm$10.3 &               WISE &    302.9 &   \citet{wright2010} \\
        4.60 &          165.2$\pm$5.5 &               WISE &    168.1 &   \citet{wright2010} \\
        6.00 &         104.1$\pm$10.2 &                IRS &    101.6 &            this work \\
        7.00 &           80.2$\pm$7.0 &                IRS &     75.5 &            this work \\
        7.97 &           67.7$\pm$4.5 &                IRS &     58.7 &            this work \\
        9.00 &           58.6$\pm$3.6 &                IRS &     46.4 &            this work \\
        9.00 &           68.7$\pm$7.0 &                IRC &     46.4 & \citet{ishihara2010} \\
       10.00 &           52.8$\pm$3.4 &                IRS &     37.8 &            this work \\
       10.99 &           50.4$\pm$2.6 &                IRS &     31.3 &            this work \\
       11.56 &           49.1$\pm$2.3 &               WISE &     28.4 &   \citet{wright2010} \\
       13.05 &           50.8$\pm$2.8 &                IRS &     22.3 &            this work \\
       14.83 &           56.2$\pm$4.9 &                IRS &     17.3 &            this work \\
       16.99 &           70.4$\pm$6.5 &                IRS &     13.3 &            this work \\
       19.02 &           88.5$\pm$7.7 &                IRS &     10.6 &            this work \\
       21.28 &         115.2$\pm$15.6 &                IRS &      8.5 &            this work \\
       22.09 &          161.5$\pm$9.7 &               WISE &      7.9 &   \citet{wright2010} \\
       23.67 &          153.1$\pm$3.1 &               MIPS &      6.8 &     \citet{chen2012} \\
       24.48 &         160.8$\pm$15.2 &                IRS &      6.4 &            this work \\
       25.00 &         186.0$\pm$33.5 &               IRAS &      6.1 &       \citet{moshir} \\
       27.45 &         207.5$\pm$19.7 &                IRS &      5.1 &            this work \\
       30.50 &         249.5$\pm$19.1 &                IRS &      4.1 &            this work \\
       33.55 &         309.7$\pm$23.4 &                IRS &      3.4 &            this work \\
       60.00 &         684.0$\pm$61.6 &               IRAS &      1.1 &       \citet{moshir} \\
   70.00$^b$ &         738.7$\pm$52.5 &               PACS &      0.8 &            this work \\
       71.42 &         659.2$\pm$33.4 &               MIPS &      0.7 &     \citet{chen2012} \\
  100.00$^b$ &         637.0$\pm$45.5 &               PACS &     0.37 &            this work \\
      160.00 &         382.3$\pm$27.9 &               PACS &     0.14 &            this work \\
      250.00 &         156.4$\pm$11.5 &              SPIRE &     0.06 &            this work \\
      350.00 &           84.3$\pm$8.3 &              SPIRE &     0.03 &            this work \\
      500.00 &           35.4$\pm$8.9 &              SPIRE &     0.01 &            this work \\
      870.00 &            8.5$\pm$4.4 &             LABOCA &    0.005 &  \citet{nilsson2010} \\
\enddata                                                                        
\tablenotetext{a}{The quoted flux densities are not color corrected.}           
\tablenotetext{b}{The emission is marginally resolved at these wavelengths.}    
\end{deluxetable}


\begin{deluxetable}{lcccccc}
\tabletypesize{\scriptsize}
\tablecaption{Properties of host stars in currently known gaseous debris systems. 
\label{gaseousdisks1}}
\tablewidth{0pt}                                                                                       
\tablecolumns{7}
\tablehead{\colhead{ID} & \colhead{D} & \colhead{T$_{\rm eff}$} & \colhead{L$_*$} 
& \colhead{M$_*$} & \colhead{Age} & \colhead{Group} \\
\colhead{} & \colhead{(pc)} &  \colhead{(K)} & \colhead{(L$_{\odot}$)} & \colhead{(M$_{\odot}$)} & 
\colhead{(Myr)} & \colhead{}  
}
\startdata
49\,Ceti    & 59.4 &  8900 (5) & 16.4 (5) & 2.00 (5) & 40$\pm10$ (3,11) & ARG   \\
HD\,21997   & 71.9 &  8300 (5) & 11.2 (5) & 1.85 (5) & 30$\pm10$ (3,11) & COL   \\
$\beta$\,Pic&19.44 &  8200 (2) &  8.7 (2) & 1.75 (2) & 23$\pm4$ (4) & BPMG  \\
HD\,131835  & 122.7&  8250 (10) &  9.2 (10) & 1.77 (10) & 16$\pm1$ (7) & UCL   \\
HD\,32297   & 112.4&  8000 (9) &  6.2 (9) & 1.65 (9) & 15--500 (9)   & - \\
HD\,172555  & 28.5 &  7800 (7) &  7.8 (7) & 1.68 (7) & 23$\pm4$ (4) & BPMG  \\
$\eta$\,Tel & 48.2 &  9500 (1) & 20.9 (1) & 2.20 (1) & 23$\pm4$ (4) & BPMG  \\
\enddata 
\tablecomments{Col.(1): Identification. Col.(2): Distance. Based on {\sl Hipparcos} trigonometric parallaxes. 
Col.(3): Effective temperature. Col.(4): Luminosity. Col.(5): Stellar mass. Col.(6): Stellar age.
For members of young moving groups and associations                                                                                                                              
 we adopted the age of the corresponding group.
Col.(7): Membership status of the star. For group names see Table~\ref{basictable} caption. 
References for the literature data used in this table: 1) \citet{chen2014}; 2) \citet{crifo1997}; 3) \citet{gagne2014}; 4) \citet{mamajek2014};
5) \citet{moor2015}; 6) \citet{pecaut2011}; 7) \citet{rm2012}; 8) \citet{roberge2013}; 9) \citet{rodigas2014}; 10) this work; 11) \citet{torres2008}. }  
\end{deluxetable} 

\begin{deluxetable}{lcccccc}
\tabletypesize{\scriptsize}
\tablecaption{Fundamental disk properties of known gaseous debris systems around A-type stars. \label{gaseousdisks2}}
\tablewidth{0pt}                                                                                       
\tablecolumns{7}
\tablehead{\colhead{ID} & \colhead{L$_{\rm IR}$/L$_{\rm bol}$} & \colhead{M$_{\rm dust}$}  & \colhead{Gas detection} & \colhead{M$_{\rm CO}$} & \colhead{Variable} & \colhead{Proposed} \\
\colhead{} & \colhead{} & \colhead{} &
\colhead{CO/\ion{O}{1}/\ion{C}{2}/\ion{C}{1}} & \colhead{} & \colhead{abs. lines} & \colhead{origin} \\
\colhead{} & \colhead{} & \colhead{(M$_{\oplus}$)} & \colhead{} & \colhead{(M$_{\oplus}$)} & \colhead{} & \colhead{}
}
\startdata
49\,Ceti     & 1.1$\times$10$^{-3}$ (14) &  2.7$\pm$0.4$\times$10$^{-1}$ (14)  &  Y/Y/Y/Y (22,19,20)  & $>$2.5$\times$10$^{-4}$ (4) & Y (12) & Secondary (19) \\
HD\,21997    & 5.7$\times$10$^{-4}$ (14) &  1.6$\pm$0.5$\times$10$^{-1}$ (14)  &  Y/N/N/N (13)        & 4--8$\times$10$^{-2}$ (10)   & N	 & Primordial (10) \\
$\beta$\,Pic & 2.6$\times$10$^{-3}$ (15) &  7.8$\pm$0.8$\times$10$^{-2}$ (5)  &  Y/Y/Y/Y (5,18,2)    &  2.9$\times$10$^{-5}$ (5)   & Y (1) & Secondary (7,5)  \\
HD\,131835   & 3.0$\times$10$^{-3}$ (21) &  4.7$\pm$1.6$\times$10$^{-1}$ (21)  &  Y/N/N/N (21)        &$>$5.2$\times$10$^{-4}$ (21)  & N	 & ?	       \\
HD\,32297    & 4.4$\times$10$^{-3}$ (3)  &  3.7$\pm$1.1$\times$10$^{-1}$ (11)  &  N/N/Y/N (6)         & $<$2.2$\times$10$^{-4}$ (13) & N	 & Secondary (6)  \\   
HD\,172555   & 7.3$\times$10$^{-4}$ (16) &  4.8$\pm$0.06$\times$10$^{-4}$ (16) &  N/N/N/Y (16)        & $<$2.4$\times$10$^{-5}$ (13) & Y (9) & Secondary (8,16)  \\
$\eta$\,Tel  & 2.4$\times$10$^{-4}$ (3)  &  1.3$\pm$0.08$\times$10$^{-2}$ (17) &  N/N/Y/N (17)        & $<$6.8$\times$10$^{-5}$ (21) & N	 & Secondary (17)  \\
\enddata 
\tablecomments{Col.(1): Identification. Col.(2): Fractional luminosity of the disks.
Col.(3): Dust mass. In the case of HD\,32297 the mass was estimated from the 1.3\,mm data published by \citet{meeus2012}, using 
the outline described in Sect.~\ref{dustprops}. Col. (4): Detection of CO, \ion{O}{1}, \ion{C}{2}, and \ion{C}{1} gas. 
 Col.(5): Mass of CO gas. For HD\,21997 and $\beta$\,Pic the mass estimates were taken from the literature or 
 from Table~\ref{basictable}. For the sake of homogeneity, in the case of 49\,Ceti, HD\,32297, and HD\,172555 the previous mass estimates 
 were recomputed using their CO (3--2) line flux data taken from the literature and assuming optically thin emission and 
 $T_{ex} = 20$\,K. Because of unknown $^{12}$CO optical depth the
 mass estimates for 49\,Ceti and HD\,131835 are lower limits.  
 Col.(6): Detection of variable optical absorption lines.  Col.(7): Proposed origin of the observed gas.
 References for the literature data used in this table: 1) \citet{beust1994}; 2) \citet{cataldi2014}; 
3) \citet{chen2014}; 4) \citet{dent2005}; 5) \citet{dent2014}; 6) \citet{donaldson2013}; 7) \citet{fernandez2006}; 
8) \citet{johnson2012}; 9) \citet{kiefer2014}; 10) \citet{kospal2013}; 11) \citet{meeus2012}; 12) \citet{montgomery2012}; 
13) \citet{moor2011b}; 14) \citet{moor2015}; 15) \citet{rhee2007}; 16) \citet{rm2012}; 17) \citet{rm2014};
18) \citet{roberge2006}; 19) \citet{roberge2013}; 20) \citet{roberge2014}; 21) this work; 22) \citet{zuckerman1995}.
}  
\end{deluxetable}

\end{document}